\documentclass[%
 reprint,
superscriptaddress,
 amsmath,
 amssymb,
 prl,
 aps,
]{revtex4-2}

\usepackage{CJK}

\usepackage[braket]{qcircuit}

\usepackage{enumitem}   
\usepackage{subcaption}
\usepackage{hyperref}
\usepackage{caption}

\usepackage{multirow} 
\usepackage{braket}
\usepackage{graphicx}
\usepackage{dcolumn}
\usepackage{bm}
\usepackage[version=4]{mhchem} 
\usepackage{adjustbox} 

\begin{document}

\begin{abstract}
One limitation of the variational quantum eigensolver algorithm is the large number of measurement steps required to estimate different terms in the Hamiltonian of interest. Unitary partitioning reduces this overhead by transforming the problem Hamiltonian into one containing fewer terms. We explore two different circuit constructions of the transformation required - one built by a sequence of rotations and the other a linear combination of unitaries (LCU). To assess performance, we simulated chemical Hamiltonians and studied the ground states of \ce{H2} and \ce{LiH}. Both implementations are successful even in the presence of noise. The sequence of rotations realization offers the greatest benefit to calculations, whereas the probabilistic nature of LCU reduces its effectiveness. To our knowledge, this work also demonstrates the first experimental implementation of LCU on quantum hardware.
\end{abstract}

\begin{CJK*}{GB}{}

\title{Implementation of Measurement Reduction for \\ the Variational Quantum Eigensolver}

\author{Alexis Ralli}
\email{alexis.ralli.18@ucl.ac.uk}
\affiliation{
Centre for Computational Science, Department of Chemistry, University College London, WC1H 0AJ, United Kingdom
}

\author{Peter J. Love}
\email{peter.love@tufts.edu}
\affiliation{
 Department of Physics and Astronomy, Tufts University, Medford, MA 02155, USA
}
\affiliation{Computational Science Initiative, Brookhaven National Laboratory, Upton, NY 11973, USA}

\author{Andrew Tranter}
\email{tufts@atranter.net}
\affiliation{
 Department of Physics and Astronomy, Tufts University, Medford, MA 02155, USA
}
\affiliation{Cambridge Quantum Computing, 9a Bridge Street Cambridge, CB2 1UB, United Kingdom
}

\author{Peter V. Coveney}
\email{p.v.coveney@ucl.ac.uk}
\affiliation{
Centre for Computational Science, Department of Chemistry, University College London, WC1H 0AJ, United Kingdom
}
\affiliation{
Informatics Institute, University of Amsterdam, Amsterdam, 1098 XH, Netherlands
}

\date{\today}
\maketitle
\end{CJK*}

\section{\label{sec:Intro}Introduction}
Current quantum computing devices have significant limitations, namely short coherence times, low qubit numbers and little to no error correction. These machines are termed noisy intermediate-scale quantum (NISQ) computers \cite{preskill2018quantum}. The leading candidate algorithms for use on NISQ devices are variational hybrid quantum-classical algorithms such as the variational quantum eigensolver (VQE) and quantum approximate optimization algorithm (QAOA)~\cite{farhi2014quantum, peruzzo2014variational}. VQE estimates Hamiltonian eigenvalues on near term quantum computers~\cite{peruzzo2014variational}. Many different implementations of the algorithm have been performed utilizing a wide array of different quantum platforms \cite{colless2018computation, gentile2018witness, hempel2018quantum, google2020hartree}.

VQE has been widely applied to the electronic structure problem. The second quantized form of the molecular electronic  Hamiltonian is converted to a qubit Hamiltonian by the Jordan-Wigner, Bravyi-Kitaev or related transformations \cite{seeley2012bravyi,BRAVYI2002210,jordan1993paulische}. The resulting qubit Hamiltonian is a linear combination of $m$ Pauli operators on $n$ qubits:
\begin{equation}
\label{eq:qubit_H}
    H_{q} = \sum_{i=0}^{m-1} c_{i} P_{i} =  \sum_{i=0}^{m-1} c_{i} \bigg( \sigma_{0}^{i} \otimes  \sigma_{1}^{i} \otimes ,...,\otimes  \sigma_{n-1}^{i} \bigg),
\end{equation}
where $c_{i}$ are real coefficients, $P_{i}$ are $n$-qubit Pauli operators, which are $n$-fold tensor products of $1$-qubit Pauli operators, or the $2\times2$ identity matrix: $\sigma_{j}^{i} \in \{X, Y, Z, \mathcal{I} \} \: \forall i$, and $j$ indexes the qubit the operator acts on.

In general, the number of terms in equation~(\ref{eq:qubit_H}) scales as $\mathcal{O}(N_{\text{orb}}^{4})$, where $N_{\text{orb}}$ is the number of orbitals~\cite{cao2019quantum}. The linearity of expectation values allows
\begin{equation}
\label{eq:qubit_H_exp_val}
   E(\vec{\theta}) = \langle  H_{q} \rangle  = \sum_{i=0}^{m-1} c_{i} \bra{\psi(\vec{\theta})}  P_{i} \ket{\psi(\vec{\theta})},
\end{equation}
where $\ket{\psi(\vec{\theta})}$ is an ansatz state produced by a parameterized quantum circuit. In conventional VQE, the expectation value of each subterm $\langle P_{i} \rangle$ is determined independently.

An estimate of each term's expectation value $\langle P_{i} \rangle$ is found by averaging over $M_{i}$ repeated measurement outcomes $\{ s^{(i)}_{j} \}_{j=1,2,.., M_{i}}$ via \cite{peruzzo2014variational,mcclean2016theory,guerreschi2017practical}:
\begin{equation}
\label{eq:estimate}
    \langle P_{i} \rangle = c_{i} \bra{\psi(\vec{\theta})} P_{i} \ket{\psi(\vec{\theta})} \approx c_{i} \bigg( \frac{1}{M_{i}} \sum_{j=1}^{M_{i}} s^{(i)}_{j} \bigg),
\end{equation}
where $s^{(i)}_{j} \in \{ -1,+1 \}$. The above expression is exact as the number of samples $M_{i} \rightarrow \infty$.

A finite number of runs is used to estimate each $\langle P_{i} \rangle$ term and thus each outcome will belong to a distribution centred around the expectation value $\bra{\psi(\vec{\theta})} P_{i} \ket{\psi(\vec{\theta})}$ with standard deviation $\epsilon_{i}$. Each estimate $\bra{\psi(\vec{\theta})} P_{i} \ket{\psi(\vec{\theta})}$ is derived from sums of random variables with finite variance. Due to the central limit theorem, they must  converge to a normal distribution~\cite{mcclean2016theory}. Because we will be comparing Hamiltonians with different numbers of terms, we will take a slightly different approach. We combine a single sample of all terms into a single-shot energy estimate $e_{j}=\sum_{i}c_{i}s_{j}^{(i)}$. The distribution of this estimate determines how many samples are required to achieve a given error on the mean. 

The optimal number of repetitions $M_{i}$ to achieve a certain precision $\epsilon$ is \cite{wecker2015progress, rubin2018application}:

\begin{equation}
    \label{eq:M_total}
M =  \sum_{i=0}^{m-1} M_{i} =  \frac{1}{\epsilon^{2}} \bigg( \sum_{i=0}^{m-1} |c_{i}| P_{i} \bigg)^{2} 
 \leq \frac{1}{\epsilon^{2}} \bigg(\sum_{i=0}^{m-1} |c_{i}|\bigg)^{2},
\end{equation}

\noindent where $M$ is the total number of measurements. Since the number of terms in equation \ref{eq:qubit_H} scales as $\mathcal{O}(N_{\text{orb}}^{4})$, the total number of measurements required will scale as $\mathcal{O}(N_{\text{orb}}^{6}/\epsilon^{2})$, where chemical accuracy is defined as $\epsilon = 1~{\rm kcal/mol}$ ($1.6$ $mHa$),  the accuracy required to match typical thermochemical experiments. Wecker \textit{et al.}  showed that to obtain energy estimates for \ce{HeH+}, \ce{BeH2} and \ce{H2O} requires $10^{8}-10^{9}$ samples to achieve an error of $1$ $mHa$ \cite{wecker2015progress}. This implies that the number of measurements required is an obstacle for experimental implementations of VQE to the number of qubits currently available on NISQ devices \cite{google2020hartree}. For example, take the experimental implementation of VQE by Hempel \textit{et al.}, which took $20$ $ms$ to perform each VQE repetition on a trapped ion quantum computer \cite{hempel2018quantum}. To obtain the ground state energy of \ce{H2}, in a minimal basis to within chemical accuracy, of order $14000$ repetitions were needed and $4.6$ minutes of averaging required. 

Various approaches have been proposed for reducing the total number of samples required by VQE ~\cite{kandala2017hardware, verteletskyi2020measurement,izmaylov2019revising, cotler2020quantum, bonet2019nearly, gokhale2019n, jena2019pauli, huggins2019efficient, gokhale2019minimizing}. In this paper, we focus on the unitary partitioning procedure independently proposed by Verteletskyi \textit{et al.} \cite{izmaylov2019unitary} and Zhao \textit{et al.} \cite{zhao2019measurement}. The main idea of this approach is to partition the qubit Hamiltonian into groups of $n$-fold Pauli operators whose linear combination is unitary. The overall operator represented by these sums can then be measured at once using additional coherent resources. In this work, we compare two different circuit realizations of unitary partitioning, as proposed in  \cite{zhao2019measurement}. 

\section{\label{sec:UP} Unitary partitioning}

The expectation value of any Hermitian operator can be obtained via a single set of single-qubit measurements, because it can be written in terms of its spectral decomposition. For example, consider the spectral decomposition of a general Hermitian operator $A$:

\begin{equation}
A = \sum_{a=0}^{d-1} \lambda_{a}  \ket{\psi_{a}}\bra{\psi_{a}},
\end{equation}

\noindent $d$ is the dimension of the space and $A$ acts on orthonormal states $\ket{\psi_{a}}$. Each  $\ket{\psi_{a}}$ is an eigenstate of the operator with corresponding eigenvalue $\lambda_{a}$. As the set of eigenvectors $\{ \ket{\psi_{a}} \}$ form an orthonormal basis there always exists a unitary $R$ that maps this basis to another: $R \ket{\psi_{a}} =  \ket{e_{a}}$ or $ \ket{\psi_{a}} =  R^{\dagger} \ket{e_{a}}$. The operator $A$ can be written in this basis:
 
 \begin{subequations} 
 \label{eq:UP_singularval}
    \begin{align}    
    A & = \sum_{a=0}^{d-1} \lambda_{a}  \ket{\psi_{a}}\bra{\psi_{a}} \\
       & = \sum_{a=1}^{d} \lambda_{a} R^{\dagger}  \ket{e_{a}}\bra{e_{a}} R \\
       &=R^{\dagger} \bigg( \sum_{a=1}^{d} \lambda_{a}  \ket{e_{a}}\bra{e_{a}} \bigg) R \\
       &=R^{\dagger} Q R. 
    \end{align}
\end{subequations} 
  
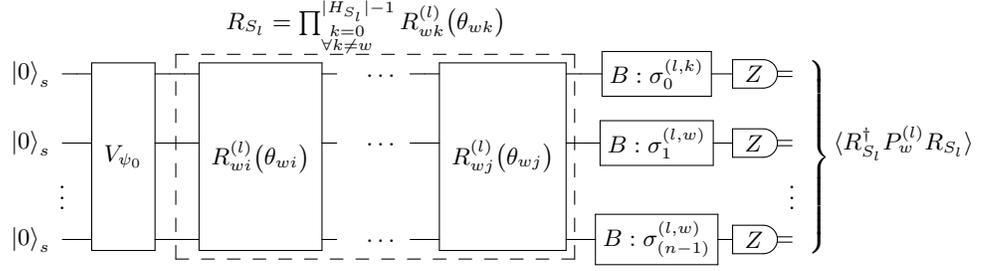
\begin{figure*}[t]
     \centering
     \hspace{9.0em}
        \Qcircuit @C=0.6em @R=1.0em {
        & & & & & & &  & & \mbox{$R_{S_{l}} =\prod_{\substack{k=0 \\  \forall k \neq w}}^{|H_{S_{l}}|-1} R_{wk}^{(l)} \big( \theta_{wk} \big)$} & & & & & & & & & & & & & & \\
        &\lstick{\ket{0}_{s}}    &\qw     &\multigate{3}{V_{\psi_{0}}}  &\qw &\qw   &\multigate{3}{R_{wi}^{(l)}\big( \theta_{wi} \big)}  &\qw  &      &  &\ldots   & & &\qw  &\multigate{3}{R_{wj}^{(l)}\big( \theta_{wj} \big)}  &\qw          &\gate{B:\sigma_{0}^{(l,k)}}     &\measureD{Z}  &\cw & & \\
        &\lstick{\ket{0}_{s}}    &\qw     &\ghost{V_{\psi_{0}}}             &\qw &\qw   &\ghost{R_{wi}^{(l)}\big( \theta_{wi} \big)}             &\qw &       & &\ldots    & &  &\qw &\ghost{R_{wj}^{(l)}\big( \theta_{wj} \big)}             &\qw       &\gate{B:\sigma_{1}^{(l,w)}}   &\measureD{Z}      &\cw  & &\rstick{\langle R_{S_{l}}^{\dagger} P_{w}^{(l)} R_{S_{l}} \rangle} \\ 
        &\vdots                        &         &                                               &       &        &                                               &       &       & &            &  &   &       &                                             &          &                                               &                             &\vdots  &  & \\ 
        &\lstick{\ket{0}_{s}}    &\qw     &\ghost{V_{\psi_{0}}}         &\qw  &\qw                 &\ghost{R_{wi}^{(l)} \big( \theta_{wi} \big)}               &\qw  &            &    &\ldots &   &  &\qw &\ghost{R_{wj}^{(l)}\big( \theta_{wj} \big)}             &\qw       &\gate{B:\sigma_{(n-1)}^{(l,w)}} &\measureD{Z}  &\cw & & 
        \relax\gategroup{2}{19}{5}{20}{1.0em}{\}}  \gategroup{2}{6}{5}{15}{0.7em}{--}
        }
\caption{General quantum circuit implementation of unitary partitioning constructed as a sequence of rotations. The subscript s denotes system qubits. $V_{\psi_{0}}$ is a unitary gate that prepares the ansatz state. To measure the qubits in the computational basis, the single qubit gates $B \in \{ H, R_{x}(-\pi/2), \mathcal{I} \}$ are required to perform a change of basis dependent on the Pauli operator $P_{w}^{(l)}$ measured.}
        \label{fig:SeqRot_UP_circ}
\end{figure*} 
 
The expectation value of $A$  can be found by $\langle A \rangle = \bra{\psi} A \ket{\psi}  = \bra{\psi} R^{\dagger} Q R \ket{\psi}$.  Note $Q$ is a matrix defined by the bracket in equation \ref{eq:UP_singularval}c. This idea underpins the unitary partitioning method. 

Individual $n$-fold tensor product operators of Pauli operators $P_{i}$ are Hermitian and unitary; however, a sum of unitary operators is in general not unitary. To make $\sum_{j} c_{j} P_{j}$ unitary three constraints are imposed \cite{izmaylov2019unitary,zhao2019measurement}: 

\begin{enumerate}
  \item $\{P_{i}, P_{j} \} = 2 \delta_{i,j}$,
  \item $\sum_{j}  |c_{j}|^{2} = 1$,
  \item $\operatorname{Im}\big( c_{j}^{*}c_{i} \big) = 0$.
\end{enumerate}

\noindent Here $\{$,$\}$ is the anticommutator ($\{ A,B \} \equiv AB+BA$). The first condition is satisfied by partitioning the qubit Hamiltonian $H_{q}$ into $m_{c}$ sets denoted $\{ S_{l} \}_{l=0,2,.., m_{c}-1}$. The sub-Hamiltonian corresponding to each anticommuting set $H_{S_{l}}$ is defined as:
\begin{equation}
\label{eq:anti_comm_Hq}
    H_{S_{l}}=\sum_{\substack{j=0 \\  P_{j} \in S_{l}}}^{|S_{l}|-1} c_{j}^{(l)} P_{j}^{(l)}.
\end{equation}
$S_{l}$ is the set of $P_{i}$ terms in $H_{S_{l}}$, where $\left\{P_{j}, P_{i}\right\}=0$ $\forall P_{j} \neq P_{i} \in S_{l}$. The process of finding such sets is discussed in \cite{jena2019pauli, izmaylov2019unitary,zhao2019measurement} and formulated as a minimum clique cover problem. This is an NP-hard problem~\cite{karp1972reducibility}; however, heuristic algorithms can provide sufficiently good approximate solutions to this problem~\cite{tranter2019ordering,izmaylov2019unitary,verteletskyi2020measurement}. Condition $(2)$ is satisfied by re-normalising each anticommuting set:
\begin{equation}
\label{eq:anti_comm_set}
    H_{S_{l}}= \gamma_{l} \sum_{\substack{j=0 \\  P_{j} \in S_{l}}}^{|S_{l}|-1} \beta_{j}^{(l)} P_{j}^{(l)},
\end{equation}
where $\sum_{j} (\beta_{j}^{(l)})^{2}=1$ and $c_{j}^{(l)} =  \gamma_{l} \beta_{j}^{(l)}$. The final condition is already satisfied, as all the coefficients $c_{i}$ in $H_{q}$ are real.

Using the unitary partitioning method, the qubit Hamiltonian is separated into $m_{c}$ sets of unitary sums:
\begin{equation}
\label{eq:paritioning_of_Q_Hamiltonian}
    H_{q} =  \sum_{i=0}^{m-1} c_{i} P_{i} = \sum_{l=0}^{m_{c}-1} H_{S_{l}} = \sum_{l=0}^{m_{c}-1} \gamma_{l} \Bigg( \sum_{\substack{j=0 \\  P_{j} \in S_{l}}}^{|S_{l}|-1} \beta_{j}^{(l)} P_{j}^{(l)} \Bigg).
\end{equation}
Each sum $\sum_{P_{j} \in S_{l}} \beta_{j}^{(l)} P_{j}^{(l)}$ can be written as $ R_{l}^{\dagger} Q_{l} R_{l}$. The expectation value of the Hamiltonian can therefore be obtained via:
\begin{equation}
\label{eq:reduced_VQE}
   \bra{\psi} H_{q} \ket{\psi} = \sum_{l=0}^{m_{c}-1} \gamma_{l} \bra{\psi} R_{l}^{\dagger} Q_{l} R_{l} \ket{\psi},
\end{equation}
and only $m_c \leq m $ terms are estimated, at the expense of needing to implement $R_{l}$ coherently within each circuit \cite{zhao2019measurement}.

In this work, the operator $R_{l}$ required by unitary partitioning is implemented by either a sequence of rotations or LCU \cite{wiebe2012hamiltonian}. These methods follow the constructions outlined in \cite{zhao2019measurement}. The relevant details for each method are summarised in the next subsections. It will be shown that the new operator $Q_{l}$ is simply a particular  $n$-fold Pauli operator that we denote as $P_{w}^{(l)}$.

%

\subsection{\label{sec:UP_SeqRot} Ordered Sequence of rotations approach}
In this section, $R_{l}$ is constructed such that it maps a completely anticommuting set  $H_{S_{l}}$ (equation \ref{eq:anti_comm_Hq}) to a single Pauli operator via conjugation - formally: $R_{l}  \bigg( \frac{H_{S_{l}}}{\gamma_{l}}  \bigg)  R_{l}^{\dagger} =  P_{w}^{(l)}$. To begin a particular $P_{j} \in S_{l}$ is selected to be reduced to. This term is denoted  by the index $w$ and written as $P_{w}^{(l)}$, where $l$ indexes the set. Once chosen, this operator is used to define the following set of Hermitian self-inverse operators \cite{zhao2019measurement}:

\begin{equation}
 \label{eqn:X_sk}
 \{ \mathcal{X}_{wk}^{(l)} = iP_{w}^{(l)}P_{k}^{(l)} \: | \: \forall P_{k} \in S_{l} \: where \:  k\neq w  \},
\end{equation}

\noindent note the coefficients $\beta_{w}$ and $\beta_{k}$ are not present. As every $P_{j}$ operator in ${S_{l}}$ anticommutes with all other operators in the set by definition, it is clear from equation \ref{eqn:X_sk} that $\mathcal{X}_{wk}^{(l)} $ will commute with all $\{P_{j} |  \: \forall P_{j} \in S_{l} \: where \: j\neq w,k \}$ and anticommute with $P_{w}^{(l)}$ and $P_{k}^{(l)}$. This property is the crux of this conjugation approach.

The adjoint rotation generated by $\mathcal{X}_{wk}^{(l)} $  can be written \cite{zhao2019measurement}:

\begin{equation}
 \label{eqn:R_sk}
	R_{wk}^{(l)} = e^{ \big( -i \frac{\theta_{wk}}{2}\mathcal{X}_{wk}^{(l)} \big)},
\end{equation}

\noindent whose action on $H_{S_{l}}$ is \cite{zhao2019measurement}:

\begin{equation}
    \label{eqn:R_sk_action}
\begin{aligned}
R_{wk}^{(l)} \bigg( \frac{H_{S_{l}}}{\gamma_{l}}  \bigg) R_{wk} ^{\dagger (l)} =& \; \big( \beta_{k} \cos\theta_{wk} - \beta_{w} \sin\theta_{wk} \big) P_{k}^{(l)} \\ 
&+ \big( \beta_{k} \sin\theta_{wk} + \beta_{w} \cos\theta_{wk} \big) P_{w}^{(l)} \\ 
&+ \sum_{\substack{P_{j} \in S_{l} \\ \forall j \neq w,k}} \beta_{j} P_{j}.
\end{aligned}
\end{equation}

\noindent The coefficient of $P_{k}^{(l)}$ can be made to go to $0$, by setting $ \beta_{k} \cos\theta_{wk} = \beta_{w} \sin\theta_{wk}$. This approach removes the term with index $k$ and increases the coefficient of $P_{w}^{(l)}$ from $\beta_{w} \mapsto\sqrt{\beta_{w}^{2} + \beta_{k}^{2}}$. This process is repeated over all indices excluding $k = w$ until only the $P_{w}^{(l)}$ term remains. This procedure can be concisely written using the following operator:

\begin{equation}
 \label{eqn:R_S}
	R_{S_{l}}  =\prod_{\substack{k=0 \\  \forall k \neq w}}^{|H_{S_{l}}|-1} R_{wk}^{(l)} \big( \theta_{wk} \big),
\end{equation}

\noindent which is simply a sequence of rotations. The angle $\theta_{wk}$  is defined iteratively at each step of the removal process, as  the coefficient  of $P_{w}^{(l)}$ increases at each step and thus must be taken into account. Importantly the correct solution for $\theta_{wk}$ must be chosen given the signs of $\beta_{w}$ and $\beta_{k}$ \cite{zhao2019measurement}. The overall action of this sequence of rotations is:

\begin{equation}
 \label{eqn:RsHsRs}
	R_{S_{l}}  \bigg( \frac{H_{S_{l}}}{\gamma_{l}}  \bigg)  R_{S_{l}} ^{\dagger} =  P_{w}^{(l)}.
\end{equation}

\noindent Fig. \ref{fig:SeqRot_UP_circ} shows the general circuit for unitary partitioning implementation as a sequence of rotations.

Applying  $R_{S_{l}}$ on $ \bigg( \frac{H_{S_{l}}}{\gamma_{l}}  \bigg)$ by conjugation maps the unitary sum $H_{S_{l}}$ to a single Pauli operator $P_{w}^{(l)}$ and unitary partitioning has been achieved. To summarise, in order to measure the expectation value of the Hamiltonian the following set of measurements are required:

\begin{equation}
    \label{eq:VQE_conj}
\begin{aligned}
H_{q} =  \sum_{i=0}^{m-1} c_{i} P_{i} = \sum_{l=0}^{m_{c}-1} H_{S_{l}} = \sum_{l=0}^{m_{c}-1} \gamma_{l} \bigg( \frac{H_{S_{l}}}{\gamma_{l}}  \bigg) \\ = \sum_{l=0}^{m_{c}-1} \gamma_{l} R_{S_{l}}^{\dagger} P_{w}^{(l)} R_{S_{l}},
\end{aligned}
\end{equation}

\noindent where the number of terms in the qubit Hamiltonian requiring separate measurement is reduced from $m \mapsto m_{c}$. Note that $R_{S_{l}} \equiv  R_{l}$, we use this notation to differentiate this assembly from the LCU construction.

\subsection{\label{sec:LCU_approach}  Linear combination of unitaries method}

\begin{figure*}[t]
\centering
  \begin{subfigure}[b]{0.9\textwidth}
		\includegraphics[scale=0.8]{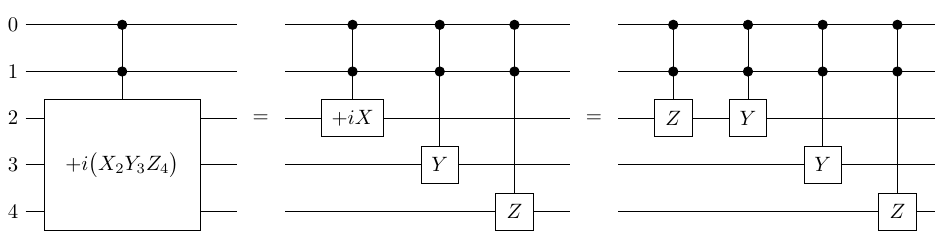}
		\caption{Example of $n$-fold $P_{i}$ with $+i$ complex phase. Note that $YZ=+iX$ giving $+i (X_{2}Y_{3}Z_{4}) = (Y_{2} Z_{2}) Y_{3}Z_{4}$. Note other possibilities are equally valid: $+i (X_{2}Y_{3}Z_{4}) = X_{2} (Z_{3}  X_{3}) Z_{4} = X_{2} Y_{3} (X_{4}  Y_{4})$ }\label{fig:pos_phase_case}	
  \end{subfigure}
  \begin{subfigure}[b]{0.9\textwidth}
		\includegraphics[scale=0.8]{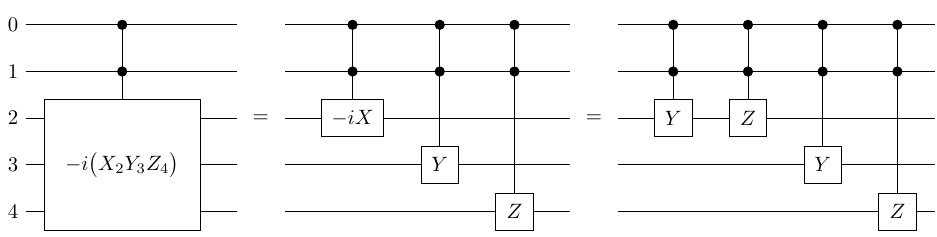}
		\caption{Example of $n$-fold $P_{i}$ with $-i$ complex phase. Note that $ZY=-iX$ giving $-i (X_{2}Y_{3}Z_{4}) = (Z_{2} Y_{2}) Y_{3}Z_{4}$. Note other possibilities are equally valid: $-i (X_{2}Y_{3}Z_{4}) = X_{2} (X_{3}  Z_{3}) Z_{4} = X_{2} Y_{3} (Y_{4}  X_{4})$}\label{fig:neg_phase_case}
  \end{subfigure}
  \begin{subfigure}[b]{0.9\textwidth}
		\includegraphics[scale=0.7]{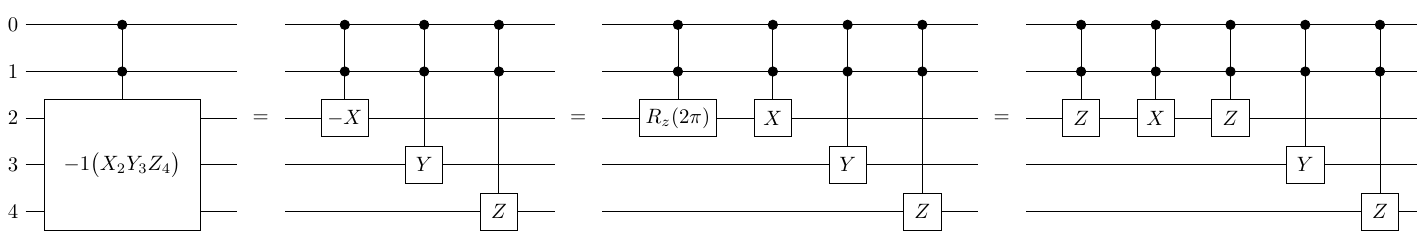}
		\caption{Example of $n$-fold $P_{i}$ with $-1$ complex phase. Note that $ \sigma^{i} R_{z}(2\pi) =R_{z}(2\pi) \sigma^{i}= \sigma^{j}\sigma^{i}\sigma^{i}= - \sigma^{i}$ $\forall i \neq j$, where $\sigma^{i} \in \{X,Y,Z \}$. This allows $-1 (X_{2}Y_{3}Z_{4}) = (X_{2} R_{z}[2\pi]) Y_{3}Z_{4} = (Z_{2} X_{2}  Z_{2}) Y_{3}Z_{4}$. Note other possibilities are equally valid: $-1 (X_{2}Y_{3}Z_{4}) =X_{2} (Y_{3}  R_{z}[2\pi] )Z_{4} = X_{2}Y_{3} (Z_{4}  R_{z}[2\pi]) = (Y_{2} X_{2}  Y_{2}) Y_{3}Z_{4}  = X_{2} (X_{3} Y_{3}  X_{3}) Z_{4} = X_{2} (Z_{3} Y_{3}  Z_{3}) Z_{4} = X_{2} Y_{3} (X_{4} Z_{4}  X_{4})= X_{2} Y_{3} (Y_{4} Z_{4}  Y_{4})$. }\label{fig:neg_real_case}
  \end{subfigure}
	\caption{Implementation of $n$-fold tensor products of Pauli operators with a complex phase of (a) $+i$, (b) $-i$ and (c) $-1$.}\label{fig:P_word_phase}
\end{figure*}

\subsubsection{\label{sec:LCU_method} LCU technique overview}

Given a complex operator as a linear combination of $d$ unitary operators:
\begin{equation}
\label{eq:LCU_A_op}
    A=\sum_{j=0}^{d-1} \alpha_{j} U_{j}, \: \: \: \|A\| \leq\|\vec{\alpha}\|_{1}=\sum_{j=1}^{d}|\alpha_{j}|
\end{equation}

\noindent where $U_{j}$ are unitary operators (that are assumed to be easy-to-implement) and $\alpha_{j}$ are real positive coefficients. Without loss of generality phase factors and signs can be absorbed into the unitaries $U_{j}$ to make all $\alpha_{j} \geq 0$. Fig. \ref{fig:P_word_phase} shows how to do this for $n$-fold Pauli operators. The linear combination of unitaries (LCU) method offers a way to probabilistically implement such an operator using the two unitary operators $G$ and $U_{LCU}$ \cite{wiebe2012hamiltonian, Low2019hamiltonian}:

\begin{center}
\begin{equation}
\label{eq:LCU_U}
    U_{LCU}=\sum_{j=0}^{d-1}\ket{j}_{a} \bra{j}_{a} \otimes U_{j},
\end{equation}
\end{center}

\begin{center}
\begin{equation}
\label{eq:LCU_G}
    G=\sum_{j=0}^{d-1} \sqrt{\frac{\alpha_{j}}{\|\vec{\alpha}\|_{1}}} \ket{j}_{a} \bra{0}_{a} + \ldots
\end{equation}
\end{center}

\noindent where subscript $a$ denotes the ancilla register. $U_{LCU}$ is sometimes known as the ``select'' operator and $G$ the  
``prepare'' operator. The most important property of the unitary operator $G$ is that the coefficients $\alpha_{j}$ only define the first column of the matrix - resulting in  $G\ket{0}_{a} \mapsto \ket{G}_{a}$. The rest of the columns must be orthogonal, but can have any values - hence there is a freedom of choice when defining $G$. A practical note on this is if one finds a quantum circuit that performs $\ket{0}_{a} \mapsto \ket{G}_{a}$, then its action on other basis states will automatically be accounted for and $G$ is completely defined (provided the quantum circuit is composed as a product of unitaries). To summarise the LCU method, first $G$ is used to initialize the ancilla register: $G\ket{0}_{a} \mapsto \ket{G}_{a}$. The controlled unitary $U_{LCU}$ is then applied across the system and ancilla registers, resulting in  \cite{Low2019hamiltonian}:

\begin{center}
\begin{equation}
\begin{aligned}
    U_{LCU} \ket{G}_{a}\ket{\psi}_{s} \mapsto & \ket{G}_{a} \frac{A}{\|\vec{\alpha}\|_{1}} \ket{\psi}_{s} + \\
    &\ket{G^{\perp}}_{a} \sqrt{1- \|\frac{A}{\|\vec{\alpha}\|_{1}} \ket{\psi}_{s}\|^{2}}\ket{\psi}_{s}.
\end{aligned}
\end{equation}
\end{center}

\noindent If $\ket{G}_{a}$ is measured in the ancilla register, then the state will be projected onto $\frac{A}{\|\vec{\alpha}\|_{1}} \ket{\psi}_{s}$ and $A$ was successfully applied to the system state $\ket{\psi}_{s}$. If any other state in the ancilla register is measured (orthogonal complement $\in \mathcal{H}_{G^{\perp}}$), then the quantum state is projected into the wrong part of the Hilbert space and $\frac{A}{\|\vec{\alpha}\|_{1}} \ket{\psi}_{s}$ is not performed.

The LCU method gives a probabilistic implementation of the matrix $A$, which has a probability of success given by:

\begin{center}
\begin{equation}
\label{eq:LCU_prob}
\begin{aligned}
    P_{\text{success}}  &=   \bra{\psi}_{s} \bra{G}_{a}  \frac{A^{\dagger}}{\|\vec{\alpha}\|_{1}} \frac{A}{\|\vec{\alpha}\|_{1}}  \ket{G}_{a} \ket{\psi}_{s} \\
    &=\big( \frac{1}{\|\vec{\alpha}\|_{1}}^{2} \big) \bra{\psi}_{s}  A^{\dagger} A  \ket{\psi}_{s}.
    \end{aligned}
\end{equation}
\end{center}

As $A$ can be a non-unitary matrix $A^{\dagger} A $ doesn't necessarily result in the identity matrix. In general the success probably therefore depends on both $\ket{\psi}$ and $\big( \frac{1}{\|\vec{\alpha}\|_{1}}^{2} \big)$. For the special case when $A$ is a unitary matrix then the success probability is given by $\big( \frac{1}{\|\vec{\alpha}\|_{1}}^{2} \big)$, due to $A^{\dagger} A  = \mathcal{I}$ .

To increase the probability of success different techniques such as oblivious amplitude amplification \cite{OblivAmp14, guerreschi2019repeat} and amplitude amplification \cite{grover1997quantum, boyer1998tight} can be used. However, extra coherent resources are required.

An alternate way to describe the LCU method is to note that the matrix $A$ is in general not unitary. To build a unitary form, the normalised matrix $A$ can be embedded into a larger Hilbert space by putting it into the upper-left block of a unitary matrix \cite{gilyen2019quantum}:

\begin{center}
\begin{equation}
\begin{aligned}
    U &=
    \begin{bmatrix}
    \frac{A}{\|\vec{\alpha}\|_{1}} & .  \\
    . & . 
  \end{bmatrix} \\ 
  \implies A  &= \|\vec{\alpha}\|_{1} \Bigg( \big(\bra{0}_{a} \otimes \mathcal{I}_{s} \big) U  \big( \mathcal{I}_{s}  \otimes \ket{0}_{a} \big) \Bigg),
  \end{aligned}
\end{equation}
\end{center}

\noindent where each $[.]$ represents a matrix of arbitrary elements. The reason to divide by $\|\vec{\alpha}\|_{1}$ is unitary matrices must have eigenvalues in the form $e^{i\theta}$, ensuring a norm of one. If for example $A=1000 Y$, putting this directly in the top-left block of $U$ would result in $\| U \| \neq 1$. Whereas, embedding $A/1000= Y$ ensures that $U$ can have a norm of one ($\| U \| = 1$) -  now dependent on remaining $[.]$ parts.

The operator $U$ is a probabilistic implementation of $A$ and is commonly known as a `block encoding' of $A$. Overall the LCU method encodes the desired matrix  as  \cite{Low2019hamiltonian}:

\begin{center}
\begin{equation}
    \bra{G}_{a} U_{LCU} \ket{G}_{a} = \frac{A}{\|\vec{\alpha}\|_{1}}.
\end{equation}
\end{center}

A final note on notation. When a matrix e.g. $A$ is block encoded using $U$ we usually say  ``the unitary $U$ gives a $(\alpha, k, \epsilon)$-block encoding of $A$". Here $\alpha$ is the $l_{1}$-norm of the matrix to be block encoded. $k$ is the extra ancilla qubits required to perform the block encoding. This depends on the number of operators in the linear combination of unitaries (equation \ref{eq:LCU_A_op}) and scales logarithmically as $k=\lceil log_{2}(|A|) \rceil$, where $|A|$ is the number of operators in the linear combination.

Finally, $\epsilon$ is the error of the block encoding and is determined by \cite{gilyen2019quantum}:

\begin{center}
\begin{equation}
    \epsilon = \Bigg|\Bigg|A -   \|\vec{\alpha}\|_{1} \bigg( \big(\bra{0}_{a} \otimes \mathcal{I}_{s} \big) U  \big( \mathcal{I}_{s}  \otimes \ket{0}_{a} \big) \bigg) \Bigg|\Bigg|.
\end{equation}
\end{center}

The LCU technique can be used to implement non-unitary operations \cite{wiebe2012hamiltonian, subramanian2019implementing}, such as matrix inversion. This is achieved by constructing the required operator as a linear combination of unitaries. As the $n$-fold tensor product of Pauli operators including the $n$-fold identity operation form a 
complete operator basis, any $(2^{n} \times 2^{n})$ complex operator can be built by different linear combinations of these unitary operators.

A toy example of the LCU method is given in the next section to illustrate the practical implementation of this technique. This can be skipped without loss of continuity.

\subsubsection{\label{sec:Toy_LCU_example}Toy LCU example}

Consider the Hamiltonian $H=\alpha_{0} U_{0} + \alpha_{1} U_{1}$, where $\alpha_{0,1} \geq 0$ and  $\|\vec{\alpha}\|_{1}=|\alpha_{0}|+|\alpha_{1}|$. To implement this operator as a linear combination of unitaries $G$ (equation \ref{eq:LCU_G}) and $U_{LCU}$ (equation \ref{eq:LCU_U}) must be defined. The construction of these operators is given by the definition of $H$. In this case, they will be:

\begin{center}
\begin{equation}
    U_{LCU} = \big( \ket{0}_{a} \bra{0}_{a} \otimes U_{0} \big) + \big( \ket{1}_{a} \bra{1}_{a} \otimes U_{1} \big),
\end{equation}
\end{center}

\begin{center}
\begin{equation}
\begin{aligned}
\begin{split}
    G =& \sqrt{\frac{\alpha_{0}}{\|\vec{\alpha}\|_{1}}} \ket{0}_{a} \bra{0}_{a} + \sqrt{\frac{\alpha_{1}}{\|\vec{\alpha}\|_{1}}} \ket{1}_{a} \bra{0}_{a} \\
    & x \ket{0}_{a} \bra{1}_{a} + y \ket{1}_{a} \bra{1}_{a},\\
     =& \begin{bmatrix}
    \sqrt{\frac{\alpha_{0}}{\|\vec{\alpha}\|_{1}}} & x  \\
    \sqrt{\frac{\alpha_{1}}{\|\vec{\alpha}\|_{1}}} & y
  \end{bmatrix}.
    \end{split}
\end{aligned}
\end{equation}
\end{center}

\noindent The values of $x$ and $y$ can be anything that ensures the columns of $G$ are orthogonal. The following can be used:

\begin{center}
\begin{equation}
    G = \begin{bmatrix}
    \sqrt{\frac{\alpha_{0}}{\|\vec{\alpha}\|_{1}}} & -\sqrt{\frac{\alpha_{1}}{\|\vec{\alpha}\|_{1}}}  \\
    \sqrt{\frac{\alpha_{1}}{\|\vec{\alpha}\|_{1}}} & \sqrt{\frac{\alpha_{0}}{\|\vec{\alpha}\|_{1}}}
  \end{bmatrix},
\end{equation}
\end{center}

\noindent as discussed in \cite{subramanian2019implementing}. The quantum circuit given in Fig. \ref{fig:LCU_example} shows the probabilistic implementation of $H$ using the LCU method. Stepping through this circuit we find:

\begin{figure}[t]
\centering
\includegraphics[scale=1]{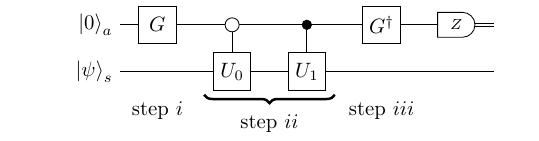}
\caption{Quantum circuit to implement $H=\alpha_{0} U_{0} + \alpha_{1} U_{1}$ as a linear combination of unitaries.}
\label{fig:LCU_example}
\end{figure}

\begin{enumerate}[label=(\roman*)]
  \item $\ket{0}_{a} \ket{\psi}_{s} \overset{G \otimes \mathcal{I}_{s}}{\mapsto} \frac{1}{\sqrt{\|\vec{\alpha}\|_{1}}} \big(\sqrt{\alpha_{0}}\ket{0}_{a} + \sqrt{\alpha_{1}}\ket{1}_{a} \big) \ket{\psi}_{s}, $
  \item   \begin{align*}
\overset{\ket{0}_{a} \bra{0}_{a} \otimes U_{0}}{\mapsto} \frac{1}{\sqrt{\|\vec{\alpha}\|_{1}}} \big( &\sqrt{\alpha_{0}}\ket{0}_{a} U_{0}\ket{\psi}_{s} \\ 
& + \sqrt{\alpha_{1}}\ket{1}_{a}\ket{\psi}_{s} \big),
\end{align*} \\
\begin{align*}
\overset{\ket{1}_{a} \bra{1}_{a} \otimes U_{1}}{\mapsto} \frac{1}{\sqrt{\|\vec{\alpha}\|_{1}}} \big( & \sqrt{\alpha_{0}}\ket{0}_{a} U_{0}\ket{\psi}_{s}  \\ 
&+ \sqrt{\alpha_{1}}\ket{1}_{a} U_{1}\ket{\psi}_{s} \big),
\end{align*}
  \item \begin{align*}
\overset{G^{\dagger} \otimes \mathcal{I}_{s}}{\mapsto} \frac{1}{\|\vec{\alpha}\|_{1}} \bigg( & \ket{0}_{a} \big( \alpha_{0} U_{0} + \alpha_{1} U_{1} \big)\ket{\psi}_{s} \\
 & + \sqrt{\alpha_{0}\alpha_{1}} \ket{1}_{a} \big( U_{1} - U_{0} \big)\ket{\psi}_{s} \bigg).
\end{align*}
\end{enumerate}

\noindent Measuring $\ket{0}_{a}$ on the ancilla line will project the state onto $\big( \alpha_{0} U_{0} + \alpha_{1} U_{1} \big)\ket{\psi}_{s} = H \ket{\psi}_{s}$ up to a normalization and heralds a successful implementation of the linear combination of unitaries method. If  $\ket{1}_{a}$ is measured then the state is projected into the wrong subspace and $H$ is not performed. Overall this approach gives a $(\|\vec{\alpha}\|_{1},1,0)$-block encoding of $H$.

If $H$ is a unitary operator, then the success probability (equation \ref{eq:LCU_prob})  is $\big( \frac{1}{\|\vec{\alpha}\|_{1}}\big)^{2}$. However, if $H$ is non-unitary then the success probability depends on $\big( \frac{1}{\|\vec{\alpha}\|_{1}} \big)^{2}  \bra{\psi}_{s} \big( \alpha_{0} U_{0}^{\dagger} + \alpha_{1} U_{1}^{\dagger} \big) \big( \alpha_{0} U_{0} + \alpha_{1} U_{1} \big) \ket{\psi}_{s} $. Consider the case of $H = \alpha_{0}U_{0} + \alpha_{1}U_{1}= \frac{1}{2} \mathcal{I} + \frac{1}{2} Z = \ket{0} \bra{0}$. This matrix defines the projector onto the all zero state on the system register's qubit and is a non-unitary operation. Clearly the success probability of block encoding this operator will depend heavily on $\ket{\psi}_{s}$ (and whether it has overlap with $\ket{0}_{s}$).

\subsubsection{\label{sec:LCU_unitary_part}  LCU approach to unitary partitioning}

Analogous to the sequence of rotations method (Section \ref{sec:UP_SeqRot}), $R_{l}$ will be applied by conjugation to an anticommuting set $H_{S_{l}}$ to give a single Pauli operator. However, unlike the previous method, where $R_{l}$ was achieved by a sequence of rotations (equation \ref{eqn:R_S}) here it is built via a linear combination of unitaries (LCU). An overview of the LCU method is given in the previous subsection.

To construct $R_{l}$ via LCU, we first need to manipulate each of the  $m_{c}$ anticommuting sets ($H_{S_{l}}$) that the qubit Hamiltonian was partitioned into (equation \ref{eq:paritioning_of_Q_Hamiltonian}). As before,  a particular Pauli operator $P_{j} \in S_{l}$ in each set is selected to be reduced to. Again this will be denoted  by the index $w$ and written as $P_{w}^{(l)}$, where $l$ indexes the set. At this point constructions of $R_{l}$ diverge. To begin we define the operator $H_{S_{l} \backslash \{{P_{w}^{(l)}}\}}$:

\begin{subequations} \label{eqn:H_n_1}
    \begin{align}
      H_{S_{l} \backslash \{{P_{w}^{(l)}}\}}= \sum_{\substack{\forall k \neq w}}^{|H_{S_{l}}|-1} \delta_{k} P_{k} ,\\
      \text{where} \sum_{\substack{\forall k \neq w}}^{|H_{S_{l}}|-1} \delta_{k}^{2}=1.
    \end{align}
\end{subequations} 

\noindent Taking each normalised set $H_{S_{l}}$ (equation \ref{eq:anti_comm_set}) and re-writing them with the term we are reducing to ($\beta_{w}^{(l)} P_{w}^{(l)} $) outside the sum:

\begin{equation}
\label{eq:anti_comm_set_excluding_P_n}
    \frac{H_{S_{l}}}{\gamma_{l}}= \beta_{w}^{(l)} P_{w}^{(l)} +  \sum_{\substack{j=0 \\  \forall j \neq w}}^{|H_{S_{l}}|-2} \beta_{j}^{(l)} P_{j}^{(l)},
\end{equation}

\noindent by re-normalising the remaining sum in equation \ref{eq:anti_comm_set_excluding_P_n}:

\begin{subequations} \label{eqn:anti_comm_set_excluding_P_n_RENORMALISED}
    \begin{align}
      \frac{H_{S_{l}}}{\gamma_{l}}= \beta_{w}^{(l)} P_{w}^{(l)} +  \Omega_{l} \sum_{\substack{j=0 \\  \forall j \neq w}}^{|H_{S_{l}}|-2} \delta_{j}^{(l)} P_{j}^{(l)}, \\
      \text{where} \sum_{\substack{j=0 \\  \forall j \neq w}}^{|H_{S_{l}}|-2} \delta_{j}^{2(l)}=1, \\
      \text{and } \beta_{j}^{(l)} = \Omega_{l} \delta_{j}^{(l)}.
    \end{align}
\end{subequations} 

\noindent We can substitute equation \ref{eqn:H_n_1}a into equation \ref{eqn:anti_comm_set_excluding_P_n_RENORMALISED}a:

\begin{equation}
\label{eq:anti_comm_set_relabelled_using_H_def}
          \frac{H_{S_{l}}}{\gamma_{l}}= \beta_{w}^{(l)} P_{w}^{(l)} +  \Omega_{l} H_{S_{l} \backslash \{{P_{w}^{(l)}}\}},
\end{equation}

\noindent where $\beta_{w}^{2(l)} +  \Omega_{l}^{2} = 1$. In  this form we can use the trigonometric identity $\cos^{2}(\theta)+\sin^{2}(\theta)=1$ to define the following operator:

\begin{equation}
\label{eq:Hn}
          H_{w}^{(l)} = \cos(\phi_{w}^{(l)}) P_{w}^{(l)}  + \sin(\phi_{w}^{(l)}) H_{S_{l} \backslash \{{P_{w}^{(l)}}\}}.
\end{equation}

\noindent Comparing equations \ref{eq:anti_comm_set_relabelled_using_H_def} and \ref{eq:Hn} it is clear that $ \cos(\phi_{w}^{(l)}) =\beta_{w}^{(l)}$ or $ \sin(\phi_{w}^{(l)})= \Omega_{l}$. Next using the definition of $H_{w}^{(l)}$ in equation \ref{eq:Hn} it was shown in \cite{zhao2019measurement} that one can consider rotations of $H_{w}$  around an axis that is Hilbert-Schmidt orthogonal to both $H_{S_{l} \backslash \{{P_{w}^{(l)}}\}}$ (equation \ref{eqn:H_n_1}) and $P_{w}^{(l)}$:

\begin{equation}
 \label{eqn:X}
	\mathcal{X}^{(l)}= \frac{i}{2} \big[ H_{S_{l} \backslash \{{P_{w}^{(l)}}\}} , P_{w}^{(l)} \big] = i \sum_{\substack{k=0 \\  \forall k \neq w}}^{|H_{S_{l}}|-2} \delta_{k}^{(l)} P_{k}^{(l)} P_{w}^{(l)}.
\end{equation}

\noindent $\mathcal{X}^{(l)}$  anticommutes with $H_{w}^{(l)}$,  is self-inverse and has the following action \cite{zhao2019measurement}:

\begin{equation}
 \label{eqn:X_action_on_Hn}
	\mathcal{X}^{(l)} H_{w}^{(l)}= i \big( -\sin \phi_{w}^{(l)} P_{w}^{(l)} +  \cos \phi_{w}^{(l)} H_{S_{l} \backslash \{{P_{w}^{(l)}}\}} \big).
\end{equation}

\noindent This defines the rotation:

\begin{subequations} \label{eqn:R_LCU}
    \begin{align}
      R_{l} &= e^{ \big( -i \frac{\alpha^{(l)}}{2} \mathcal{X}^{(l)} \big)} = \cos \big( \frac{\alpha^{(l)}}{2} \big) \mathcal{I} - i \sin \big( \frac{\alpha^{(l)}}{2} \big)  \mathcal{X}^{(l)} \\
      &=  \cos \big( \frac{\alpha^{(l)}}{2} \big) \mathcal{I} - i \sin \big( \frac{\alpha^{(l)}}{2} \big) \bigg(  i \sum_{\substack{k=0 \\  \forall k \neq w}}^{|H_{S_{l}}|-2} \delta_{k}^{(l)} P_{k}^{(l)} P_{w}^{(l)} \bigg) \\
             &=  \cos \big( \frac{\alpha^{(l)}}{2} \big) \mathcal{I} + \sin \big( \frac{\alpha^{(l)}}{2} \big)  \sum_{\substack{k=0 \\  \forall k \neq w}}^{|H_{S_{l}}|-2} \delta_{k}^{(l)} P_{kw}^{(l)},
    \end{align}
\end{subequations}

\begin{figure*}[t]
 \centering
         \hspace{2.0em}
         \Qcircuit @C=0.7em @R=1.0em {
        &\lstick{\ket{0}_{s}}    &\qw     &\multigate{2}{V_{\psi_{0}}}  &\qw &\qw               &\multigate{5}{U_{LCU}^{(l)}}   &\qw                        &\qw  &\gate{B:\sigma_{0}^{(l,w)}}     &\measureD{Z}  &\cw     &\\
        &\vdots                  &        &                             &    &                  &                         &                           &     &                                &              &\vdots  &\rstick{\langle R_{l}^{\dagger} P_{w}^{(l)} R_{l} \rangle}   \\ 
        &\lstick{\ket{0}_{s}}    &\qw     &\ghost{V_{\psi_{0}}}         &\qw &\qw               &\ghost{U_{LCU}^{(l)}}          &\qw                        &\qw  &\gate{B:\sigma_{(n-1)}^{(l,w)}} &\measureD{Z}  &\cw     & \\ 
        &\lstick{\ket{0}_{a}}    &\qw     &\qw                          &\qw &\multigate{2}{G^{(l)}}  &\ghost{U_{LCU}^{(l)}}          &\multigate{2}{G^{(l) \dagger}} &\qw  &\qw                             &\measureD{Z}  &\cw     & \\ 
        &\vdots                  &        &                             &    &                  &                         &                           &     &                                &              &\vdots  & \rstick{\text{ iff } \ket{00...0}_{a}}  \\ 
        &\lstick{\ket{0}_{a}}    &\qw     &\qw                          &\qw &\ghost{G^{(l)}}          &\ghost{U_{LCU}^{(l)}}         &\ghost{G^{(l) \dagger}}        &\qw  &\qw                             &\measureD{Z}  &\cw     & \\
&    &     &  & &                 & \mbox{$\langle 00..0 |_{a} G^{(l) \dagger} U_{LCU}^{(l)} G^{(l)}| 00...0 \rangle_{a} = R_{l}/ \|\alpha^{(l)}\|_{1}$}   &                        &  &    &  &     & 
        \relax\gategroup{1}{11}{3}{12}{1.0em}{\}} 
        \relax\gategroup{4}{11}{6}{12}{1.0em}{\}}
        \relax\gategroup{1}{6}{6}{8}{1.0em}{--}
        }
        \caption{
General quantum circuit implementation of unitary partitioning constructed as a LCU.
The subscripts $s$ and $a$ denote the system and ancilla registers respectively. $V_{\psi_{0}}$ is a unitary gate that prepares the ansatz state. To measure the qubits in the computational basis, the single qubit gates$B \in \{ H, R_{x}(-\pi/2), \mathcal{I} \}$  are required to perform a change of basis dependent on the Pauli operator $P_{w}^{(l)}$ measured.}
        \label{fig:LCU_circuit_UP}
\end{figure*}
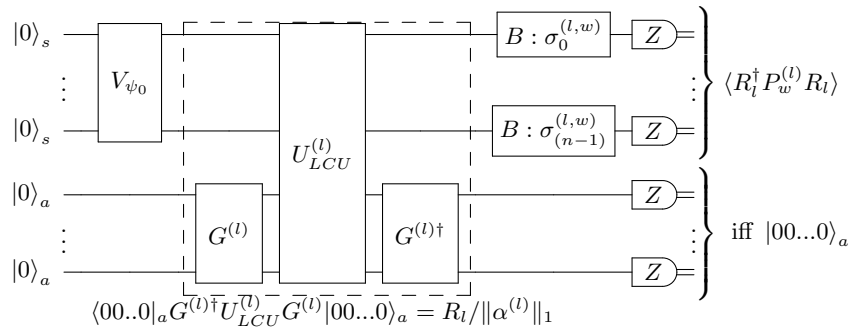

\begin{figure}[t]
     \centering
    \begin{subfigure}[t]{0.32\textwidth}
        \raisebox{-\height}{\includegraphics[width=\textwidth]{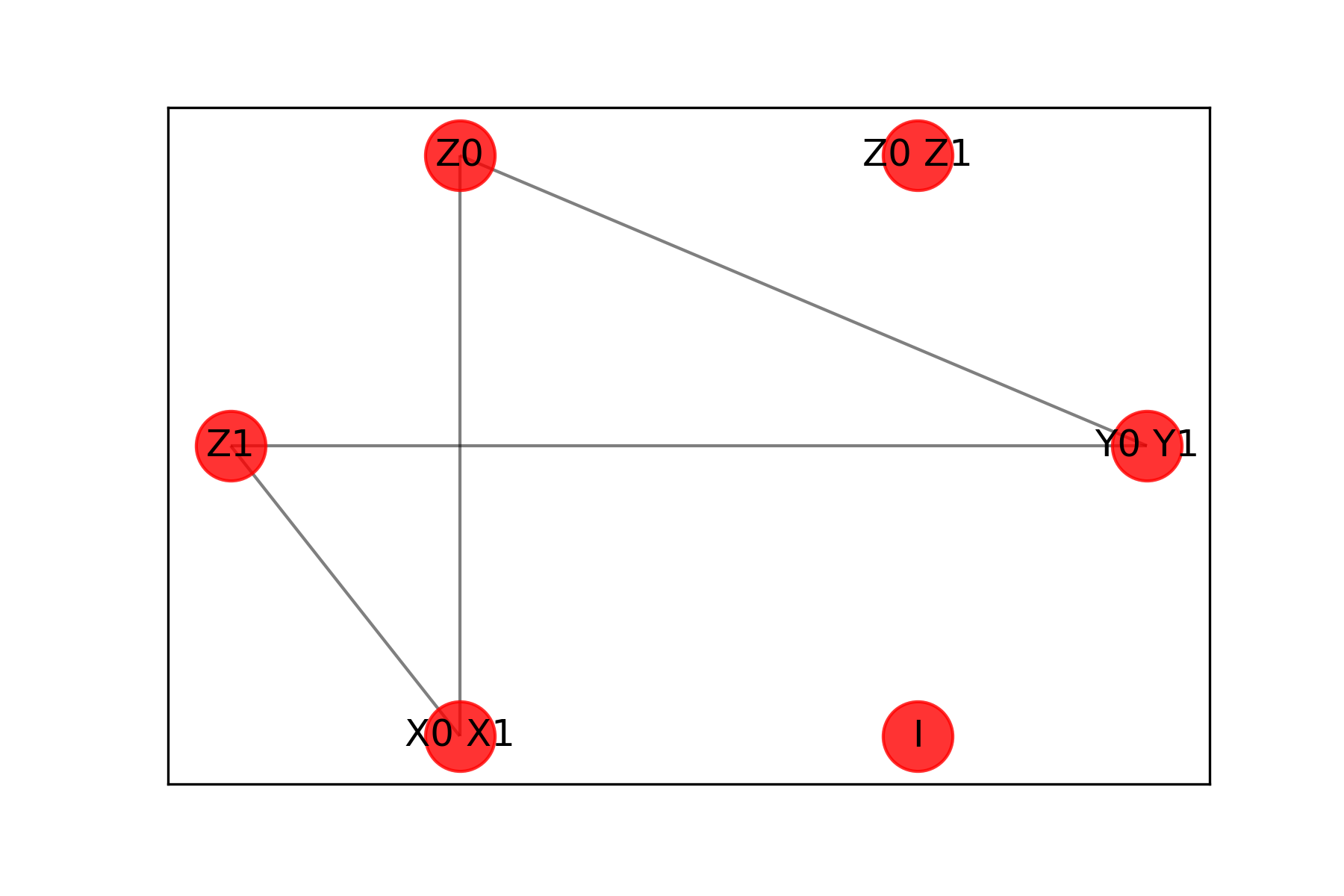}}
        \caption{Graph of qubit Hamiltonian of \ce{H2}. Edges connect anticommuting $P_{i}$ nodes.}
    \end{subfigure}
    \hfill
    \begin{subfigure}[t]{0.32\textwidth}
        \raisebox{-\height}{\includegraphics[width=\textwidth]{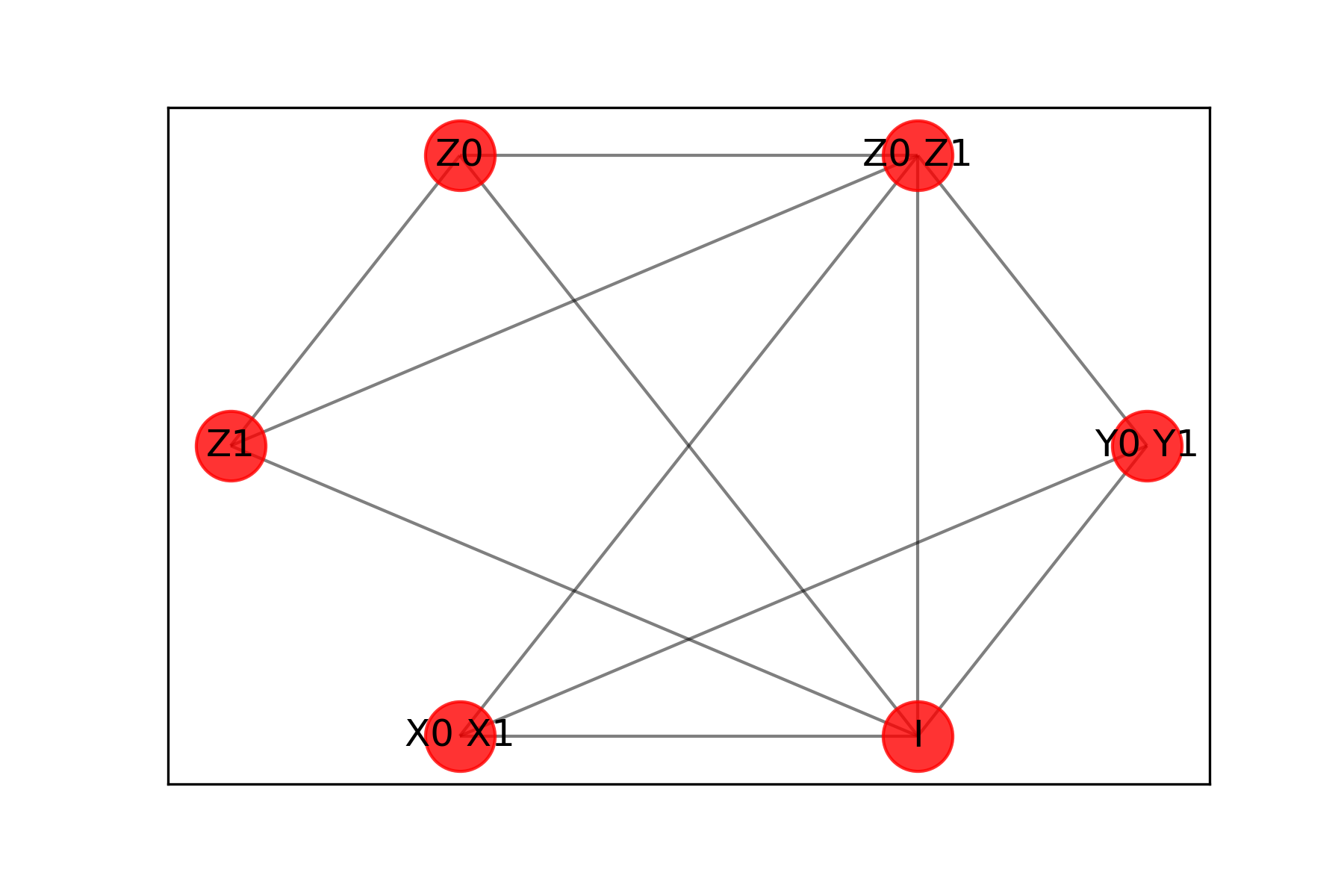}}
        \caption{Complementary graph of (a)}
    \end{subfigure}
        \hfill
    \begin{subfigure}[t]{0.32\textwidth}
        \raisebox{-\height}{\includegraphics[width=\textwidth]{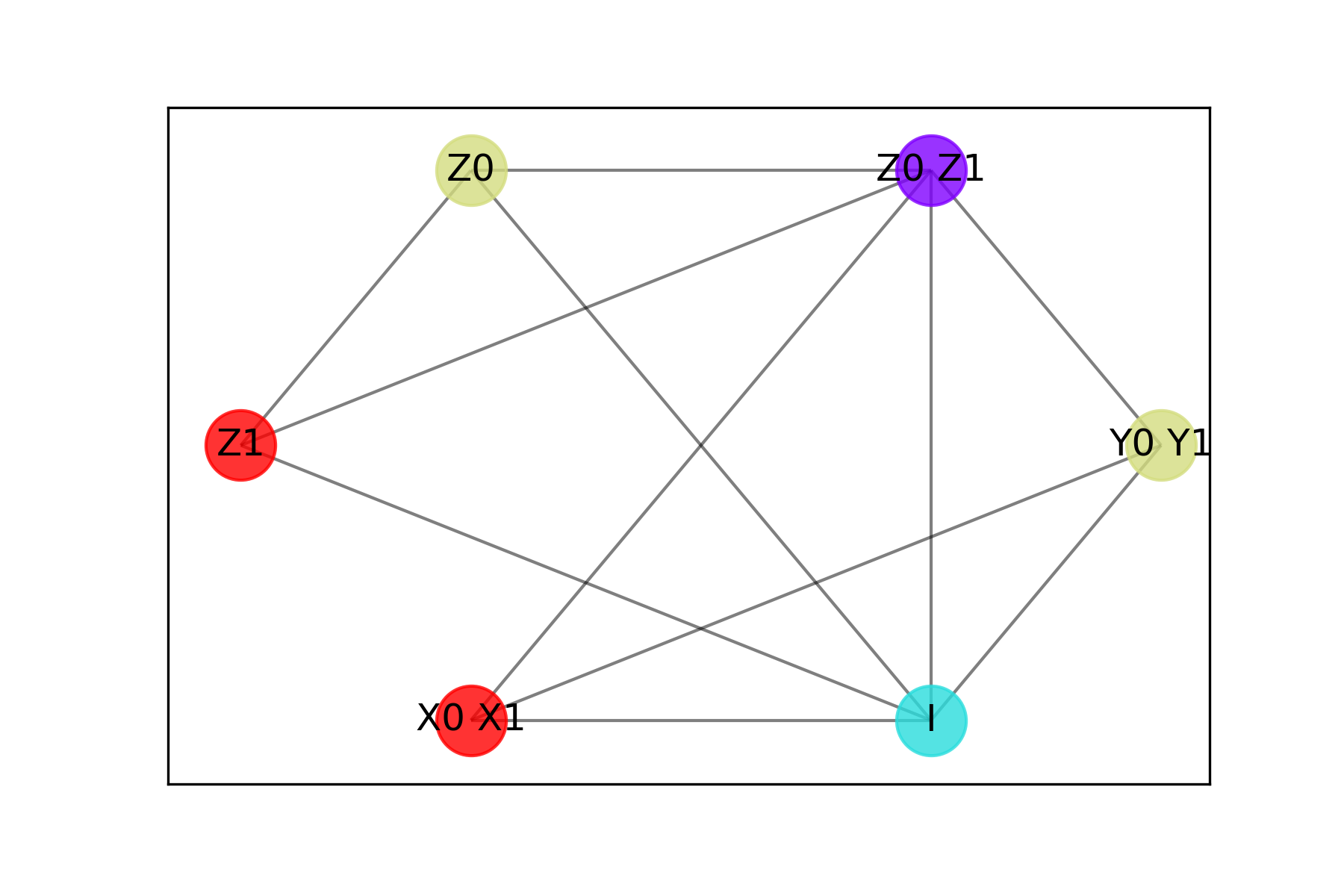}}
        \caption{Clique cover of (a) found by graph colouring of complementary graph (b)}
    \end{subfigure}

    \caption{Illustration of graph colouring approach to finding anticommuting sets of a given Hamiltonian.}
    \label{fig:H2_Graph}
\end{figure}

\begin{figure*}[t]
     \centering
    \begin{subfigure}[t]{0.32\textwidth}
        \raisebox{-\height}{\includegraphics[width=\textwidth]{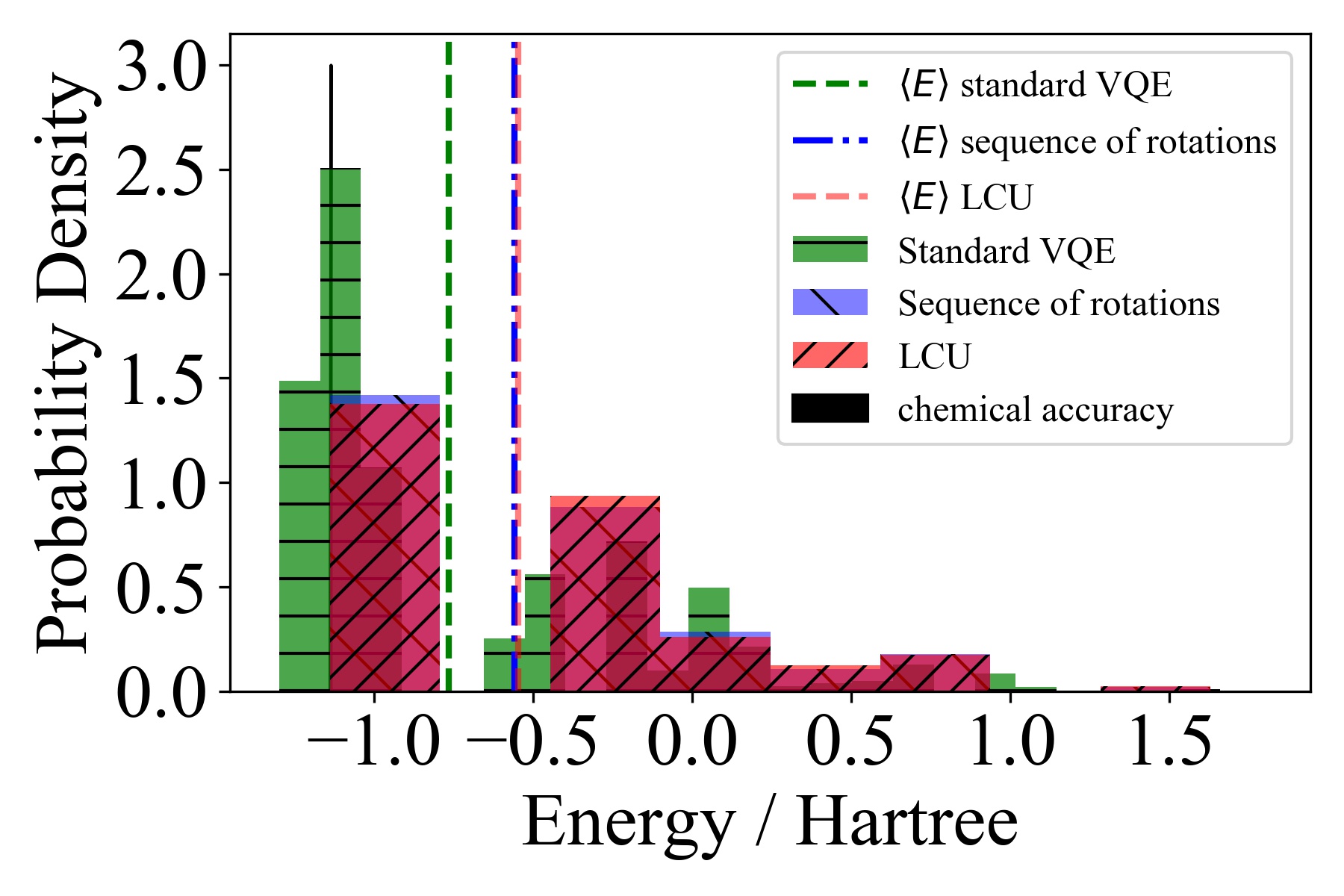}}
        \caption{}
    \end{subfigure}
    \hfill
    \begin{subfigure}[t]{0.32\textwidth}
        \raisebox{-\height}{\includegraphics[width=\textwidth]{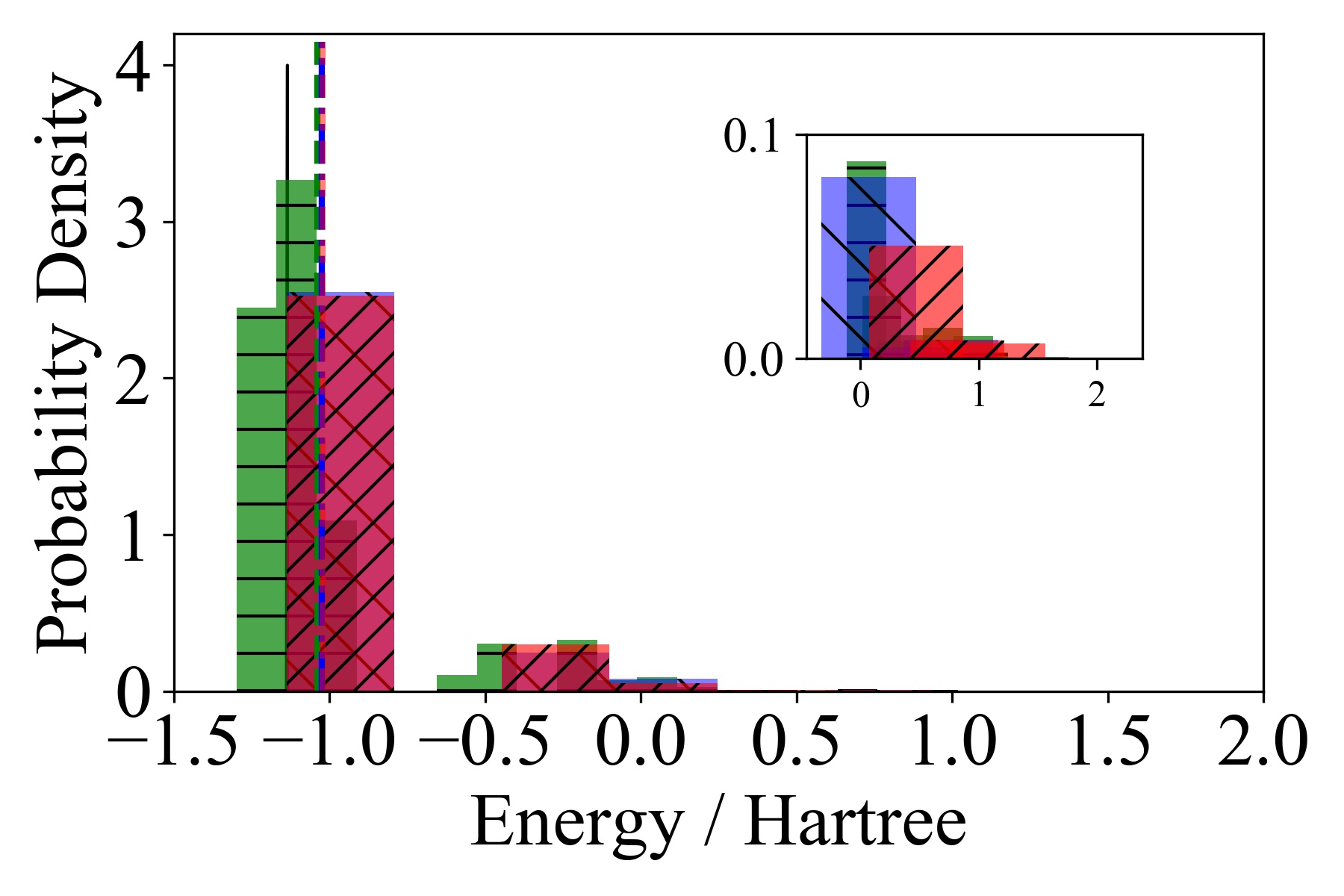}}
        \caption{}
    \end{subfigure}
    \hfill
    \begin{subfigure}[t]{0.32\textwidth}
        \raisebox{-\height}{\includegraphics[width=\textwidth]{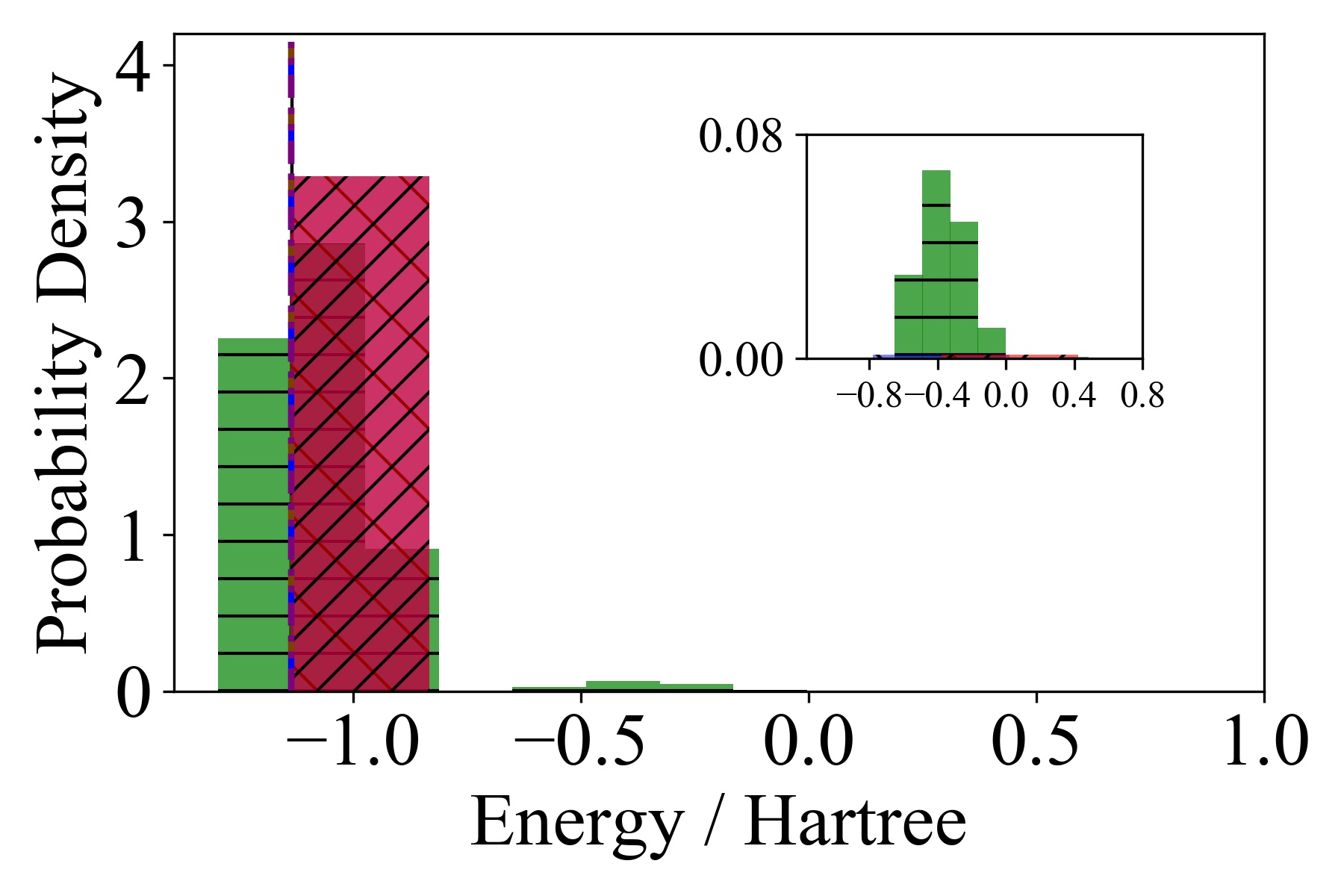}}
        \caption{}
    \end{subfigure}
    \caption{Probability density functions of single-shot VQE estimates of the ground state energy $e_{j}$ of \ce{H2}. A bin is given to every possible energy outcome. Note $E_{FCI}=-1.13728$ $Ha$, $t_{o}=5, t_{p}=3$ and $M=1.2663 \times 10^{6}$. The raw results from ibmqx2 are given in (a), (b) shows these results with measurement error mitigation applied and (c) gives results from simulation on a noise-free QPU.}
    \label{fig:H2_Hist_Results}
\end{figure*} 


\noindent where $P_{k}^{(l)} P_{w}^{(l)}  = P_{kw}^{(l)}$. Importantly $ P_{kw}^{(l)}$ will be another tensor product of Pauli operators, as products of $n$-fold Pauli operators will yield another operator in the Pauli group. The adjoint action of this rotation on $H_{w}^{(l)}$ is:

\begin{equation}
    \label{eqn:R_LCU_adjoint_action}
\begin{aligned}
R_{l} H_{w}^{(l)} R^{\dagger}_{l} = &  \sin \big( \phi_{w}^{(l)} - \theta^{(l)} \big) H_{S_{l} \backslash \{{P_{w}^{(l)}}\}} + \\ & \cos \big( \phi_{w}^{(l)} - \theta^{(l)} \big) P_{w}^{(l)}.
\end{aligned}
\end{equation}

\noindent By setting  $\theta^{(l)} = \phi_{w}^{(l)}$, the coefficient of $H_{S_{l} \backslash \{{P_{w}^{(l)}}\}}$ will go to zero and we achieve the intended result of $R_{l} H_{w}^{(l)} R^{\dagger}_{l} = P_{w}^{(l)}$. To build $R_{l}$ by the LCU method, we use its definition in equation \ref{eqn:R_LCU}c. In practice, it is easier to re-write equation \ref{eqn:R_LCU} using the fact that all $P_{kw}$ and $\mathcal{I}$ are in the Pauli group. The terms can thus be combined into a single sum:

\begin{subequations} \label{eqn:R_LCU_re_written}
    \begin{align}
      R_{l} =&  \alpha \mathcal{I} +  \sum_{\substack{k=0 \\  \forall k \neq w}}^{|H_{S_{l}}|-2} \alpha_{k} P_{kw}^{(l)} \\
      =&   \sum_{\substack{q=0 \\  \forall q \neq w}}^{|H_{S_{l}}|-1} \alpha_{q} P_{q}^{(l)}.
    \end{align}
\end{subequations}

\noindent Note all $\alpha_{q}$ must be real and $\alpha_{q}\geq 0$ $\forall q$. This is achieved by absorbing any signs and complex phases into $P_{kw}^{(l)}$, hence these operators are $n$-fold tensor Pauli operators up to a complex phase. When written in this form, it is easy to define the operators $G$ (equation \ref{eq:LCU_G}) and $U_{LCU}$ (equation \ref{eq:LCU_U}):

\begin{center}
\begin{equation}
\label{eq:LCU_G_unitaryP}
    G^{(l)}=\sum_{q=0}^{|H_{S_{l}}|-1} \sqrt{\frac{\alpha_{q}}{\|\vec{\alpha_{q}}\|_{1}}} \ket{q}_{a} \bra{0}_{a} + \ldots
\end{equation}
\end{center}

\begin{center}
\begin{equation}
\label{eq:LCU_U_unitaryP}
    U_{LCU}^{(l)} =\sum_{q=0}^{|H_{S_{l}}|-1} \ket{q}_{a} \bra{q}_{a} \otimes P_{q}^{(l)},
\end{equation}
\end{center}

\noindent that are required to perform $R_{l}$ as a LCU. Overall the operator is encoded as:

\begin{center}
\begin{equation}
    \bra{0}_{a}G^{(l) \dagger} U_{LCU}^{(l)} G^{(l)} \ket{0}_{a} = \bra{G^{(l)}}_{a} U_{LCU}^{(l)} \ket{G^{(l)}}_{a} = \frac{R_{l}}{\|\vec{\alpha_{q}}\|_{1}}.
\end{equation}
\end{center}

\noindent Without using amplitude amplification, as $R_{l}$ is unitary, the probability of success is given by the square of the $l_{1}$-norm of $R_{l}$. Note that the $l_{1}$-norm is defined as $\|\vec{\alpha_{q}}\|_{1}=\sum_{q=0}^{|H_{S_{l}}|-1}|\alpha_{q}|$.

\section{\label{sec:Num_Sim}Numerical Study}

The ability of the unitary partitioning measurement reduction strategy is dependent on the problem Hamiltonian. To assess the performance of each implementation, we investigate application to Hamiltonians of interest in quantum chemistry.

\subsection{\label{sec:method}Method}
We consider Hamiltonians for \ce{H2} and \ce{LiH} molecules employing the STO-3G and STO-6G basis sets respectively. These were calculated using Openfermion-PySCF and converted into the qubit Hamiltonian using the Bravyi-Kitaev transformation in OpenFermion \cite{mcclean2020openfermion, sun2018pyscf}.

Partitioning into anticommuting sets $H_{S_{l}}$ was performed using networkX \cite{SciPyProceedings_11}. First, a graph of the qubit Hamiltonian was built, where each node is a term in the Hamiltonian. Next edges are put between nodes on the graph that anticommute. Finally, a graph colouring of the complement graph was performed. This searches for the minimum number of colours required to colour the graph, where no neighbours of a node can have the same colour as the node itself. The ``largest first'' colouring strategy in NetworkX was used \cite{SciPyProceedings_11, welsh1967upper}.  Each unique colour represents an anticommuting clique. Fig. \ref{fig:H2_Graph} shows the method applied to \ce{H2}. This approach is the minimum clique cover problem mapped to a graph colouring problem. Further numerical details for each Hamiltonian can be found in Appendix \ref{sec:numerical_details_app}.

The input state $\ket{\psi(\vec{\theta})}$ for all calculations was the exact full configuration interaction (FCI) ground state, found by diagonalizing the Hamiltonian. As our aim was to investigate different implementations of unitary partitioning, this meant the ansatz optimization step in VQE was not required. 

For the simulations performed on IBM's ibmqx2 quantum processing unit (QPU), a measurement error mitigation strategy available in Qiskit was utilized and is a simple inversion procedure \cite{leymann2020bitter}. The quantum circuits required were generated by the qiskit.ignis \textit{complete\_meas\_cal} method and executed alongside each separate ibmqx2 experiment, with the maximum number of shots ($8192$). This sampling cost was not included in the number of calls to quantum device.  The \textit{CompleteMeasFitter} method in qiskit.ignis \cite{qiskit} was used to generate the calibration matrix required for measurement error mitigation \cite{qiskit, leymann2020bitter}.

We denote the number of terms in the original Hamiltonian by $t_{o}$ and $t_{p}$ for the unitary partitioned Hamiltonian. For each implementation, we fix $M$ the total number of calls to the QPU. This can be thought of as a measurement budget. The total number of $e_{j}$ samples - single shot estimates of all $n$-fold Pauli operators in the Hamiltonian - is $N_{o} = M / t_{o}$ for the original Hamiltonian and $N_{p} = M / t_{p}$ for the partitioned Hamiltonian. Clearly, because unitary partitioning reduces the number of terms in $H_{q}$, more energy samples are obtained for a fixed $M$.

\subsection{\label{sec:results}Results}

For a given preparation of the true ground state of \ce{H2}, we compare both implementations of measurement reduction by unitary partitioning against a standard VQE calculation on IBM's open access quantum device (ibmq 5 Yorktown - ibmqx2) and Qiskit's qasm simulator \cite{qiskit}. The quantum circuits required are given in Appendix \ref{sec:numerical_details_app}. Fig. \ref{fig:H2_Hist_Results} shows the distribution of single-shot energy estimates of all three techniques applied to molecular hydrogen. The average energy is given by $\langle E \rangle = \frac{1}{N}\sum_{j=0}^{N-1} e_{j}$. To compare each method the measurement budget was fixed to $M=1.2663 \times 10^{6}$. A calibration matrix method available in Qiskit was used to mitigate measurement errors and was used to amend the raw outputs from ibmqx2.

The qubit Hamiltonian for \ce{H2} has five terms, which is reduced to three by unitary partitioning, not including the identity term. The number of energy estimates $e_{j}$ obtained was $253260$ for standard VQE  and $5/3$ this for unitary partitioning by the sequence of rotations method. This is because the smaller number of terms allowed a correspondingly larger number of samples to be taken for a fixed $M$. The total number of $e_{j}$ samples from ibmqx2 for these techniques was reduced to $253074$ and $421951$ after measurement error mitigation was applied. The LCU approach to unitary partitioning is probabilistic and requires post-selection on the all zero state of the ancilla register. After post-selection, our simulation of unitary partitioning as a LCU on ibmqx2 gave $333407$ raw $e_{j}$ samples and $332763$ $e_{j}$ samples after measurement error mitigation was applied to the raw output. Our emulation of this method on a noise-free quantum processing unit (QPU) gave $336390$ $e_{j}$ samples after post-selection. The theoretical maximum possible number of samples for LCU would be the same as the sequence of rotations method if all samples obtained were successful.

The reason a normal distribution is not obtained is due to the number of terms in the qubit Hamiltonian for \ce{H2}. At most only $32$ distinct values of $e_{j}$ are possible for standard VQE and $8$ under unitary partitioning, and so we do not expect the central limit argument to apply here.

To investigate the distribution of energies obtained from each method in more detail, we simulated the larger problem of \ce{LiH} using  Qiskit's statevector simulator \cite{qiskit}. Further details are given in Appendix \ref{sec:LiH}. Fig. \ref{fig:LiH_Hist_Results} summarizes the results. Again, each data point is an energy estimate found from the weighted measurement outcomes of a single-shot VQE run. The standard qubit Hamiltonian for this problem is made up of $630$ terms and after applying unitary partitioning $102$ terms, not  including the  identity  term. The measurement budget was fixed at $M=1.018521 \times 10^{9}$. The total number of energy estimates $e_{j}$ for standard VQE, the sequence of rotations and the LCU methods after post-selection were $1616700$,  $9985500$ and $1447349$ respectively.

We performed the  Kolmogorov-Smirnov \cite{massey1951kolmogorov} and Shapiro-Wilk tests \cite{shapiro1965analysis} on the data in Fig. \ref{fig:LiH_Hist_Results} to check for normality. The P-values obtained in all cases were smaller than 0.05, and thus we could not assume a normal distribution. This may be caused by insufficient samples allowing convergence to the central limit or the problem size still being too small. To estimate the statistics of the true distribution we thus employed bootstrap resampling \cite{efron1992bootstrap}.

\subsection{\label{sec:Discussion}Discussion}
In our results, for a fixed measurement budget $M$ we obtain a set of independent identically distributed random energy samples $\{e_{1}, e_{2}, ...,e_{N} \}$. The standard deviation of this sample $\sigma_{e}$ converges to the true standard deviation a the number of samples increases. As the number of samples increases, the error on the sample mean decreases. The standard error of the mean (${\rm SEM}$) is defined as:

\begin{equation}
 \label{eqn:SEM}
	{\rm SEM} = \sigma_{e} / \sqrt{N},
\end{equation}

\noindent for $N$ energy samples. Because we are considering a fixed measurement budget $M$, the effect of unitary partitioning will be to increase the number of energy samples $N$ and hence reduce the SEM.

To benchmark each method we compare $\sigma_{e}$ and ${\rm SEM}$ of the ground state energies samples.  95\% confidence intervals (CI) were calculated using bootstrapping with $10,000$ resamples with replacement. The full statistical analysis is given in Table \ref{tab:H2_boot}.  

\begin{figure}[b]
\centering
\includegraphics[scale=0.5]{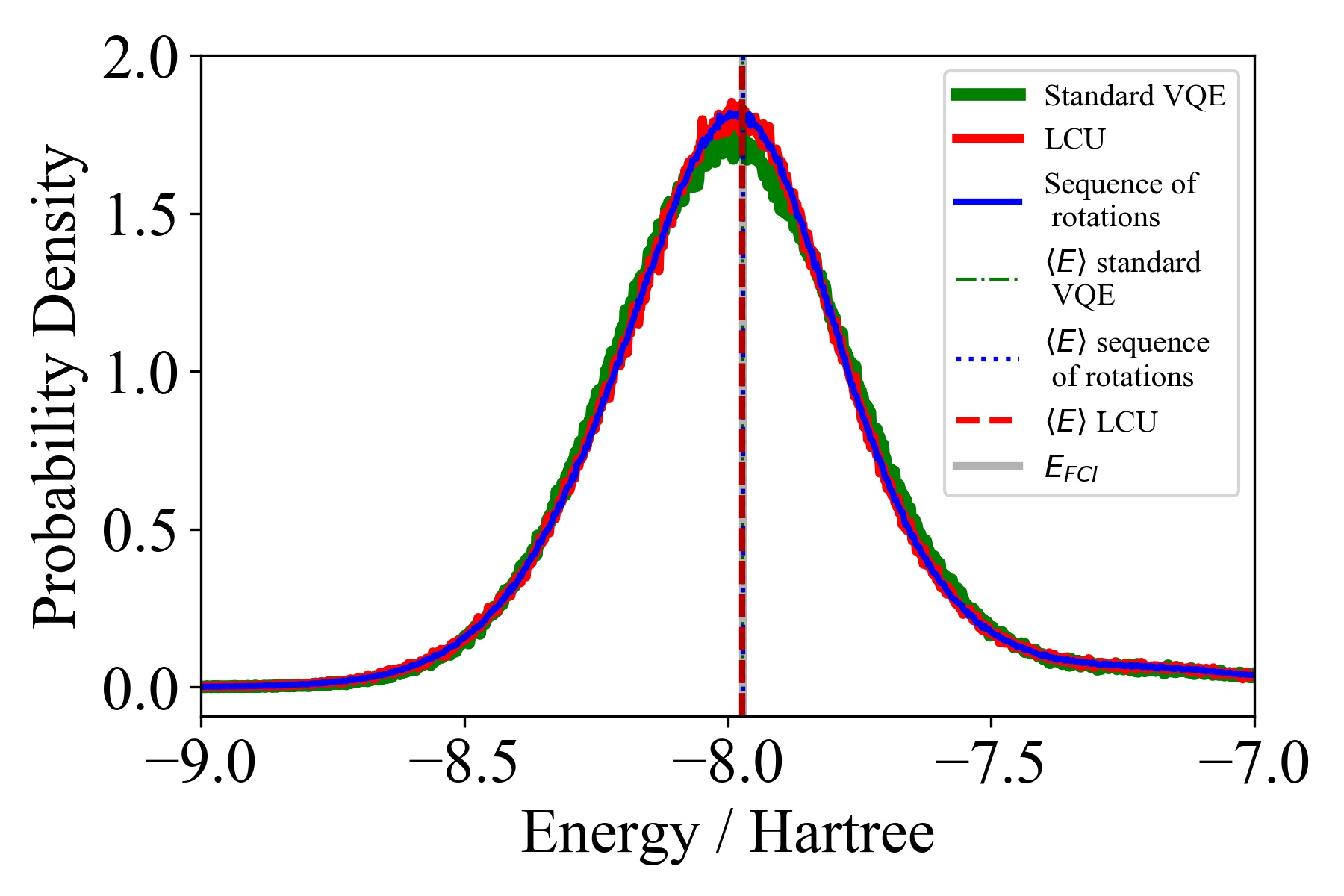}
\caption{Probability density function of single-shot VQE estimates of the ground state energy $e_{j}$ of \ce{LiH},  from a noise-free QPU simulation. The number of bins was set to $2500$ and the centre of each plotted. Note $E_{FCI}=-7.97118$ $Ha$, $t_{o}=630, t_{p}=102$ and $M=1.018521 \times 10^{9}$. Individual plots are given in Appendix \ref{sec:ind_LiH_results}. }
    \label{fig:LiH_Hist_Results}
\end{figure}

Qualitatively, the noise-free \ce{LiH} simulation results in Fig. \ref{fig:LiH_Hist_Results} show that VQE with unitary partitioning applied as either a LCU or a sequence of rotations give a similar distribution of energies compared to standard VQE.  This is expected, as unitary partitioning leaves a given molecular Hamiltonian $H_{q}$ unchanged - equation \ref{eq:paritioning_of_Q_Hamiltonian} - thus both the standard deviation $\sigma = \sqrt{ \langle H_{q} \; ^{2} \rangle - \langle H_{q} \rangle^{2}}$ and full configuration interaction ground state energy $E_{FCI}$ will remain unchanged.   

Quantitatively, the $\sigma_{e}$ of ground state energy estimates of \ce{LiH} for each method were very similar, with the largest difference being $4.8$ $mHa$. Note $\sigma$ of the full population is independent of the number of samples $N$ taken. Thus even though the distributions in  Fig. \ref{fig:LiH_Hist_Results} look very similar, the number of data points in each curve is significantly different and therefore the SEM is significantly different for each implementation.

Whereas, for the noise-free simulation of \ce{H2} (Fig. \ref{fig:H2_Hist_Results}c),  the sample standard deviation of $e_{j}$ from VQE with unitary partitioning applied were an order of magnitude lower than standard VQE. We expect this is due to the small number of distinct $e_{j}$ outcomes for this specific problem under unitary partitioning. 

As unitary partitioning is designed to require fewer terms to be measured, for a fixed measured budget $M$, the total number of energy estimates will be larger. The sequence of rotations construction of unitary partitioning is deterministic and will always give more $e_{j}$ samples than conventional VQE. Hence, the ${\rm SEM}$ of both \ce{H2} and \ce{LiH} noiseless simulations using the sequence of rotations method were an order of magnitude lower than standard VQE. On the other hand, the LCU realization of unitary partitioning is probabilistic. Even though fewer terms need measurement, post-selection requires some samples to be discarded. We see this in the simulation for \ce{LiH}, where the LCU approach actually has the fewest $e_{j}$ samples at $1447349$ compared to $1616700$ for standard VQE. As the $\sigma_{e}$ of all three approaches are similar, the LCU implementation has the highest ${\rm SEM}$ in this case. The advantage over standard VQE is thus dependent on the success probability, which for each circuit is given by the inverse $l_{1}$ norm squared of the operator to be implemented as a LCU. Importantly post-selection is only performed on the ancilla register.  The success probability is inversely proportional to the dimension of the ancilla Hilbert space. The number of ancilla qubits scales logarithmically with the number of terms in each anticommuting set ($n_{ancilla}= \lceil log_{2}(|H_{S_{l}}|-1) \rceil$), and the dimension of the ancilla Hilbert space, and hence success probability, is inversely proportional to the size of the anticommuting sets. 

The experimental results for \ce{H2} on ibmqx2 show that applying the unitary partitioning technique does not appreciably change the performance of VQE, when combined with error mitigation techniques. We suspect this is due to the extra coherent resource required to perform $R_{l}$ causing an increase in errors, which offsets the improvement of the ${\rm SEM}$ given by the technique. We expect this to be mitigated as gate fidelities increase.

\begin{table*}[t]
\centering
\begin{adjustbox}{width=1\textwidth}
\small
\begin{tabular}{cccccccccc}
\hline
Molecule &    Method &       Backend &  $N$ &         $\langle E \rangle$ / $Ha$ &                     $\langle E \rangle$ 95\%CI / $Ha$ &   $\sigma_{e}$ / $Ha$ &   $\sigma_{e}$ 95\%CI / $Ha$  &         ${\rm SEM}$ / $Ha$ &                      ${\rm SEM}$ 95\%CI / $Ha$ \\
\hline
      H2 &       LCU &  ibmqx2 - mit &     332763 &  -1.0212  &  [-1.0222, -1.0201]   &  3.1661e-01 &  [3.1504e-01, 3.1816e-01] &  5.4886e-04 &  [5.4617e-04, 5.5158e-04] \\
      H2 &    SeqRot &  ibmqx2 - mit &     421951 &  -1.0246  &  [-1.0256, -1.0237]   &  3.1984e-01 &  [3.1838e-01, 3.2132e-01] &  4.9239e-04 &  [4.9017e-04, 4.9459e-04] \\
      H2 &  standard &  ibmqx2 - mit &     253074 &  -1.0381  &  [-1.0394, -1.0367]   &  3.4240e-01 &  [3.4053e-01, 3.4424e-01] &  6.8063e-04 &  [6.7697e-04, 6.8432e-04] \\
      H2 &       LCU &  ibmqx2 - raw &     333407 &  -0.54537 &  [-0.54757, -0.54315] &  6.5287e-01 &  [6.5129e-01, 6.5442e-01] &  1.1307e-03 &  [1.1279e-03, 1.1334e-03] \\
      H2 &    SeqRot &  ibmqx2 - raw &     422100 &  -0.55863 &  [-0.56063, -0.55666] &  6.5025e-01 &  [6.4885e-01, 6.5162e-01] &  1.0009e-03 &  [9.9873e-04, 1.0030e-03] \\
      H2 &  standard &  ibmqx2 - raw &     253260 &  -0.76468 &  [-0.76685, -0.76247] &  5.6625e-01 &  [5.6441e-01, 5.6810e-01] &  1.1252e-03 &  [1.1215e-03, 1.1287e-03] \\
      H2 &       LCU &     simulator &     336390 &  -1.1373  &  [-1.1374, -1.1373]   &  1.5655e-02 &  [1.4307e-02, 1.7082e-02] &  2.6992e-05 &  [2.4692e-05, 2.9392e-05] \\
      H2 &    SeqRot &     simulator &     422100 &  -1.1373  &  [-1.1373, -1.1372]   &  1.6286e-02 &  [1.5084e-02, 1.7533e-02] &  2.5068e-05 &  [2.3147e-05, 2.6968e-05] \\
      H2 &  standard &     simulator &     253260 &  -1.1370  &  [-1.1377, -1.1363]   &  1.7752e-01 &  [1.7634e-01, 1.7872e-01] &  3.5276e-04 &  [3.5040e-04, 3.5512e-04] \\
     LiH &       LCU &  simulator &    1447349    &  -7.9719  &  [-7.9723, -7.9714]   &  2.7268e-01 &  [2.7213e-01, 2.7322e-01] &  2.2665e-04 &  [2.2620e-04, 2.2711e-04] \\
     LiH &    SeqRot &  simulator &    9985500    &  -7.9712  &  [-7.9714, -7.9710]   &  2.7292e-01 &  [2.7271e-01, 2.7312e-01] &  8.6367e-05 &  [8.6303e-05, 8.6432e-05] \\
     LiH &  standard &  simulator &    1616700.   &  -7.9716  &  [-7.9720, -7.9712]   &  2.6817e-01 &  [2.6773e-01, 2.6860e-01] &  2.1091e-04 &  [2.1056e-04, 2.1125e-04] \\
      \hline
      \end{tabular}
\end{adjustbox}
\caption{The mean, standard deviation and standard error on the mean for each method calculating the ground state energies of \ce{H2} and \ce{LiH} using single-shot VQE. The simulator backend represents a noise-free QPU emulator and ibmqx2 a real quantum device. Ibmqx2-raw are the raw experimental results from the QPU and ibmqx2-mit with measurement error mitigation applied. 95\% confidence intervals (CI) were calculated using bootstrap resampling \cite{efron1992bootstrap}. \label{tab:H2_boot}} 
\end{table*} 

The experimental execution of $R_{l}$ by LCU on ibmqx2 performed comparably to the sequence of rotations realization. Ignoring post-selection issues, the LCU algorithm is more complex and requires more qubits to implement. We believe this motivates further examination of the use of more advanced quantum algorithms on NISQ devices.

A particular feature  of our results from ibmqx2 (Fig. \ref{fig:H2_Hist_Results}) is that the mean ground state energy obtained is overestimated by a seemingly constant amount. We suspect this could be due to two effects. Firstly, our ansatz circuit prepares the ground state. Any coherent errors in this circuit will increase the energy of the state prepared by virtue of the variational principle \cite{szabo2012modern}. Secondly, inspecting the qubit Hamiltonian for \ce{H2} most coefficients are positive. As our results overestimate the energy, it implies that measurement outcomes of each $n$-fold Pauli operator are more frequently $+1$ causing each estimate of $ e_{j}$ to be larger.  This could be an indication of a higher $\ket{0}$ count on each qubit or P($0|1)>$ P($1|0$). 

\begin{figure*}[t]
\centering
\includegraphics[scale=0.8]{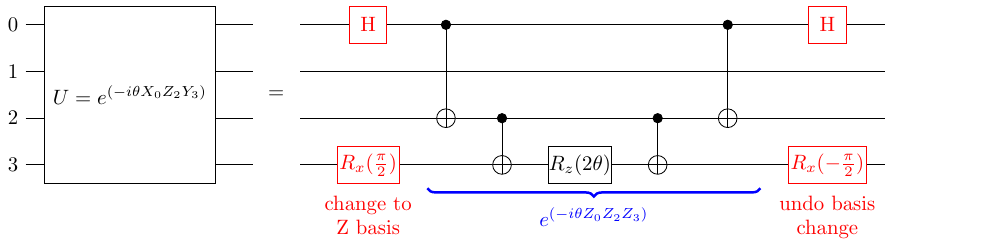}
\caption{Quantum circuit to perform a unitary operator given as an exponentiated $n$-fold tensor product of Pauli operators \cite{whitfield2011simulation}.}
\label{fig:cascade_circuit}
\end{figure*}

The single-qubit gate error rates of IBM QPU's have error rates in the range of 0.1\%-0.3\% and two-qubit gate errors in the range of 2\%-5\% \cite{tannu2019mitigating}. The most error-prone operation is measurement and ibmqx2 on average has a measurement error rate of 4\%, but this can be much higher (13\%) \cite{tannu2019mitigating}. This large measurement error is apparent when comparing the raw and measurement error mitigated results from the QPU simulation of \ce{H2}. In future experiments, it would be interesting to improve measurement fidelity, for example by using invert-and-measure designs \cite{tannu2019mitigating} as well as flipping the qubit encoding ($\ket{0} \mapsto \ket{1}$ and $\ket{1} \mapsto \ket{0}$) as in \cite{nam2020ground}, or by other mitigation schemes~\cite{chen2019detector}.

In the original work in which these techniques were proposed, it was shown that the variance of the different methods should be similar \cite{zhao2019measurement}. This is observed on both the QPU emulator and quantum device.

Crucially, when partitioning the qubit Hamiltonian into anticommuting cliques (equation \ref{eq:paritioning_of_Q_Hamiltonian}), the greatest measurement reduction is obtained if the minimum clique cover is found. This cover has the fewest $H_{S_{l}}$ sets possible. However, non-optimal clique covers still give a measurement reduction. As the size of the quantum circuit for $R_{l}$ is proportional to the number of terms in $H_{S_{l}}$, we propose that for practical applications a non-optimal clique cover is beneficial. By splitting the problem Hamiltonian into pairs of anticommuting operators ($|H_{S_{l}}|=2$ $\forall \{l\}_{l=0,1,...,m_{c}-1}$), the extra coherent resources required to perform $R_{l}$ are experimentally realistic for current and near term devices. This offers a constant factor improvement to the number of measurements required. A detailed circuit analysis of different implementations of unitary partitioning is given in the next Section.

\section{\label{sec:circ_analysis}Circuit analysis}

In order to investigate the circuit depth of each technique, we consider circuits made up of arbitrary single qubit and CNOT gates. However, often when analysing fault-tolerant protocols it is common to consider the universal gate set of Clifford and T gates. The LCU method only requires arbitrary rotations in the operator $G$, whereas the sequence of rotations method requires a rotation for every operator in an anticommuting set, apart for $P_{w}^{(l)}$. The relative depth of LCU vs sequence of rotations circuits would be interesting to explore in this setting.

\subsection{\label{sec:seq_rot_circuit} Sequence of rotations circuit analysis}

In Section \ref{sec:UP_SeqRot}, it was shown that $R_{S_{l}}$ could be constructed by a sequence of $R_{wk}^{(l)}$ rotations (equation \ref{eqn:R_S}). Writing the rotation operators in their exponentiated form :

\begin{equation}
 \label{eqn:R_S_equation}
	R_{S_{l}}  =\prod_{\substack{k=0 \\  \forall k \neq w}}^{|H_{S_{l}}|-1} R_{wk}^{(l)} \big( \theta_{wk} \big) = \prod_{\substack{k=0 \\  \forall k \neq w}}^{|H_{S_{l}}|-1}  e^{ \big( -i \frac{\theta_{wk}}{2}\mathcal{X}_{wk}^{(l)} \big)}.
\end{equation}

\noindent We need to consider the cost to perform each $R_{wk}^{(l)}$  rotation. Whitfield \textit{et al.} in \cite{whitfield2011simulation} show how to build the required quantum circuits for these types of operators and an example is illustrated in Fig. \ref{fig:cascade_circuit}.

Every $R_{wk}^{(l)}$ circuit will require $\mathcal{O} \big( 2(N_{s}-1) \big)$ CNOT gates, $1$ $R_{z}(\theta)$ gate and $\mathcal{O} (2N_{s})$ change of basis gates $\{ H, R_{x}(\theta) \}$. Here $N_{s}$ is the number of system qubits. A single $R_{wk}^{(l)}$ is needed for each term in the set $S_{l}$, apart from for $P_{w}^{(l)}$.

The total number of rotations that make up the full sequence of rotation operator $R_{S_{l}}$ is therefore $|H_{S_{l}}|-1$. $|H_{S_{l}}|$ is the size of the anticommuting set. The overall gate count scales as $\mathcal{O} \big( (2N_{s}+1) (|H_{S_{l}}|-1) \big)$ single qubit and gates and $\mathcal{O} \big( 2(N_{s}-1) (|H_{S_{l}}|-1)  \big)$ CNOT gates.

We note that there is a choice in the ordering of $R_{wk}^{(l)}$ when constructing $R_{S_{l}}$. By choosing an ordering that maximises the common substring  between Pauli strings defining $R_{wk}^{(l)}$ (lexicographical order), it is possible to cancel the common change of basis gates  between subsequent $R_{wk}^{(l)}$ rotations. We refer the reader to \cite{cowtan2019phase}, which gives further possible gate cancellations - including CNOT cancellations. This can significantly reduce the circuit depth when constructing $R_{S_{l}}$.

\subsection{\label{sec:LCU_circuit} LCU implementation}
In the linear combination of unitaries approach to unitary partitioning, $R_{l}$ is written as a linear combination of $n$-fold Pauli operators (equation  \ref{eqn:R_LCU_re_written}) - up to a complex sign. Fig. \ref{fig:P_word_phase} shows how to implement such operators.

Such an operator can be implemented using the LCU technique. To achieve this, the gates required to realise $G^{(l)}$ (equation \ref{eq:LCU_G_unitaryP}) and $U_{LCU}^{(l)}$(equation \ref{eq:LCU_U_unitaryP}) are required. We will not explicitly consider the construction of $G$, as it heavily depends on the ancilla state required and many different approaches are possible. In the worst case, without introducing any additional qubits, $\mathcal{O}(N_{c} \: log_{2}(N_{c})^{2})$  standard 1 and 2-bit gate operations are required \cite{long2001efficient}. Here the number of control qubits $N_{c}$ is given by the number of operators $U_{LCU}^{(l)}$ is constructed from. In this case $N_{c}= \lceil log_{2}(|H_{S_{l}}|-1) \rceil$.  

\begin{figure}[b]
\centering
\includegraphics[scale=0.8]{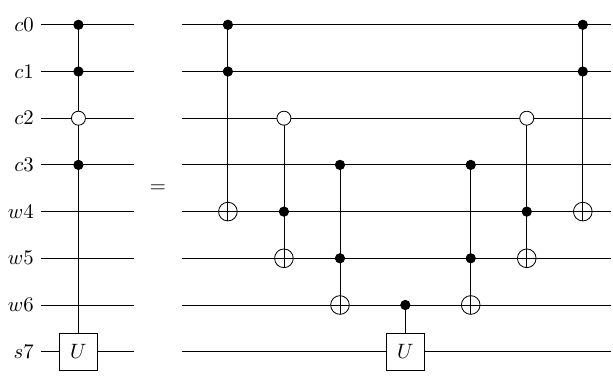}
\caption{Example quantum circuit required to perform $4$-bit controlled $U$ gate \cite{nielsen2002quantum}. Note that c, w, s denote control, work and system qubits respectively.}
\label{fig:n_contrl_gate}
\end{figure}

On the other hand, the quantum circuit to construct $U_{LCU}$ is well defined.  Overall, $(|H_{S_{l}}|-1)$ $N_{c}$-bit controlled $P_{q}$ gates are required. To efficiently construct each control $P_{q}$ gate,  $N_{w}=(N_{c}-1)$ work qubits are employed. The control states of the ancilla qubits are stored on these work qubits using Toffoli gates \cite{nielsen2002quantum}. An example is shown in Fig. \ref{fig:n_contrl_gate}. For every control $P_{i}$ a cascade of $\mathcal{O}( 2(N_{c}-1))$ Toffoli gates are required.

With no circuit simplifications on the ancilla and work qubit registers, the number of Toffoli gates required scales as $\mathcal{O} \big(2^{N_{c}} (2N_{c}-2) \big)$. However, significant simplifications can be made.  Here we assume all $2^{N_{c}}$ control states are required.

\begin{figure}[t]
\centering
  \begin{subfigure}[b]{0.4\textwidth}
    \includegraphics[scale=0.6]{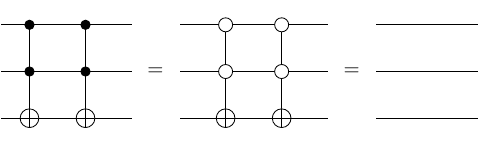}
		\caption{Gray and Binary}\label{fig:TT_1a}	
  \end{subfigure}
  \hspace{-0.75em}
  \begin{subfigure}[b]{0.4\textwidth}
    \includegraphics[scale=0.6]{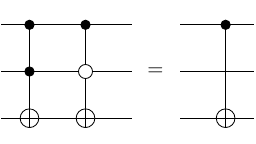}
		\caption{Gray and Binary \cite{rahman2014templates}}
		\label{fig:TT_1b}
  \end{subfigure}
  \begin{subfigure}[b]{0.4\textwidth}
    \includegraphics[scale=0.6]{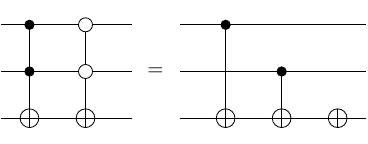}
    \caption{Binary only \cite{babbush2018encoding}}
    \label{fig:TT_1c}
  \end{subfigure}
  \begin{subfigure}[b]{0.4\textwidth}
    \includegraphics[scale=0.6]{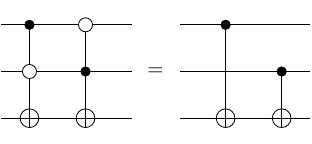}
    \caption{Binary only}
    \label{fig:TT_1d}
  \end{subfigure}
  \caption{Toffoli-Toffoli circuit templates}\label{fig:TT_gate_cancellations}
\end{figure}

\begin{figure}[b]
\centering
  \begin{subfigure}[b]{0.4\textwidth}
		\includegraphics[scale=0.6]{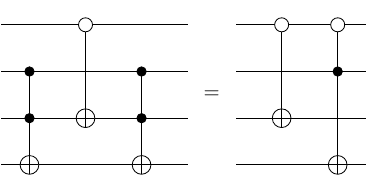}
		\caption{Gray and binary}\label{fig:TXT_1a}	
  \end{subfigure}
  \begin{subfigure}[b]{0.4\textwidth}
		\includegraphics[scale=0.6]{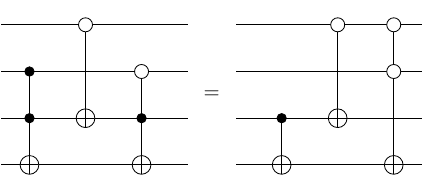}
		\caption{Binary only}\label{fig:TXT_1b}
  \end{subfigure}
  \begin{subfigure}[b]{0.4\textwidth}
		\includegraphics[scale=0.6]{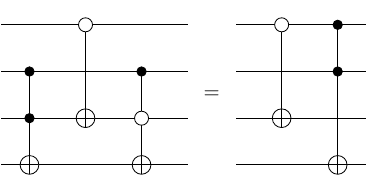}
		\caption{Not required (included for reference)}\label{fig:TXT_1c}
  \end{subfigure}
	\caption{Toffoli-CNOT-Toffoli circuit templates}\label{fig:TxT_simplifications}
\end{figure}

Importantly if we arrange the sequence of control gates optimally, we can reuse some of the work qubits when the control states overlap.  Fig. \ref{fig:TT_gate_cancellations} and Fig. \ref{fig:TxT_simplifications}  show different possible circuit templates that can be utilized to simplify the quantum circuits.  

To maximise the circuit simplifications on the ancilla and work qubit registers we show how a Gray encoding of $U_{LCU}$ should be used. Fig. \ref{fig:grey_binary} in Appendix \ref{sec:ciruit_analysis} shows the control states for $N_{c}=5$ in a Gray and binary encoding.  Importantly in a Gray code, adjacent bitstrings only differ by one bit \cite{frank1953pulse}. In other words, the Hamming distance between adjacent control states is always 1.

Consider $x$ leading bits in common between adjacent control bitstrings, where $2 \leq x \leq (N_{c}-1)$. In the circuit picture, these are the cases when the top $x$ controls between two adjacent control unitaries are in common. For these cases, we get $2(x-1)$ trivial Toffoli reductions (Fig. \ref{fig:TT_1a}) on the Toffoli gates between the control unitaries. The number of times each $x$ occurs is given by $2^{x}$ and therefore the total trivial Toffoli reduction is given by:

\begin{equation}
\begin{aligned}
\sum_{x=2}^{k} 2^{x} \; 2(x-1) &=  2^{(k+2)}(k-2) + 8 \\
     &=2^{N_{c}+1}(N_{c}-3)+8 \text{ where } k=N_{c}-1 .
\end{aligned}
\end{equation}

\begin{figure}[h!]
\centering
\includegraphics[width=0.8\textwidth]{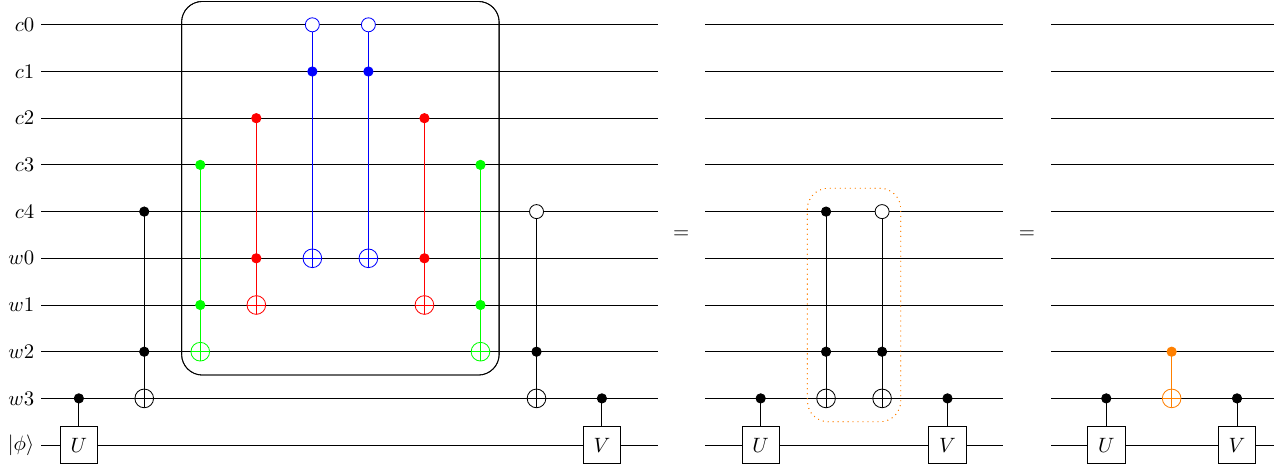}
\caption{Example of optimal circuit reduction, when pairs of adjacent control bitstrings have $N_{c}-1$ leading bits in common.  This will occur for both the Gray and binary encodings.}
\label{fig:grey_bin_full}
\end{figure}

Next, consider the case of $x=N_{c}-1$. After the trivial Toffoli gate cancellations, there must be two Toffli gates that must differ by one bit in a Gray code. These will cancel to a single CNOT - illustrated in Fig. \ref{fig:TT_1b}. This occurs $2^{x}$ times generating an additional $2^{x}= 2^{N_{c}-1} $ CNOT gates and removing a further $2^{x} (2) = 2^{x+1} = 2^{N_{c}}$ Toffoli gates. No further reductions are possible. A full example of this is given in Fig. \ref{fig:grey_bin_full}.

\begin{figure}[h!]
\centering
  \begin{subfigure}[b]{1\textwidth}
\includegraphics[scale=0.5]{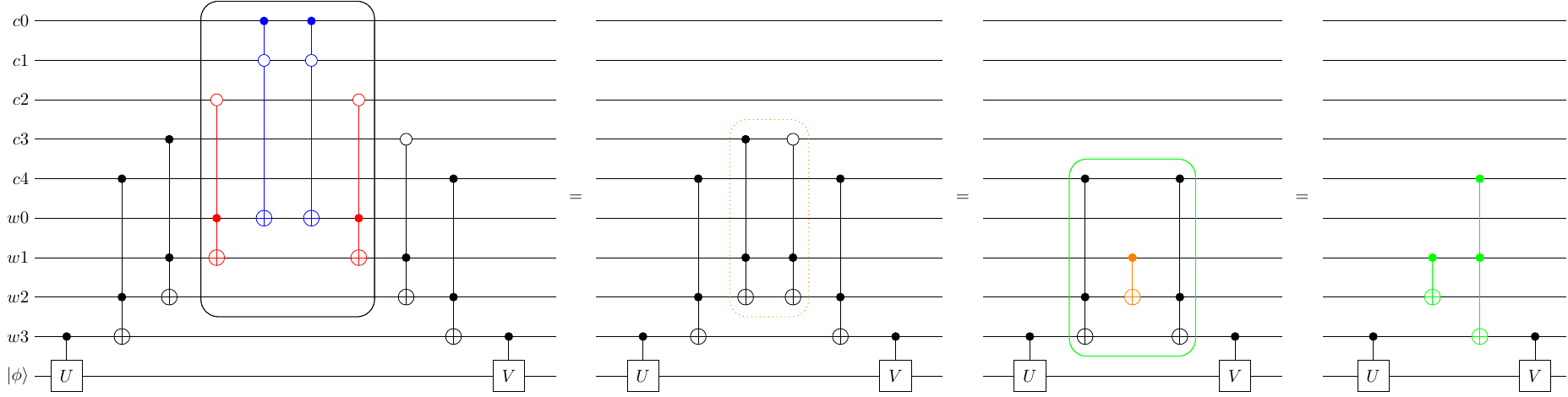}
\caption{Example reduction in Gray code for $10011\text{-}U$ and $10001\text{-}V$.}
\label{fig:grey_partial}
  \end{subfigure}
  \begin{subfigure}[b]{1\textwidth}
\includegraphics[scale=0.5]{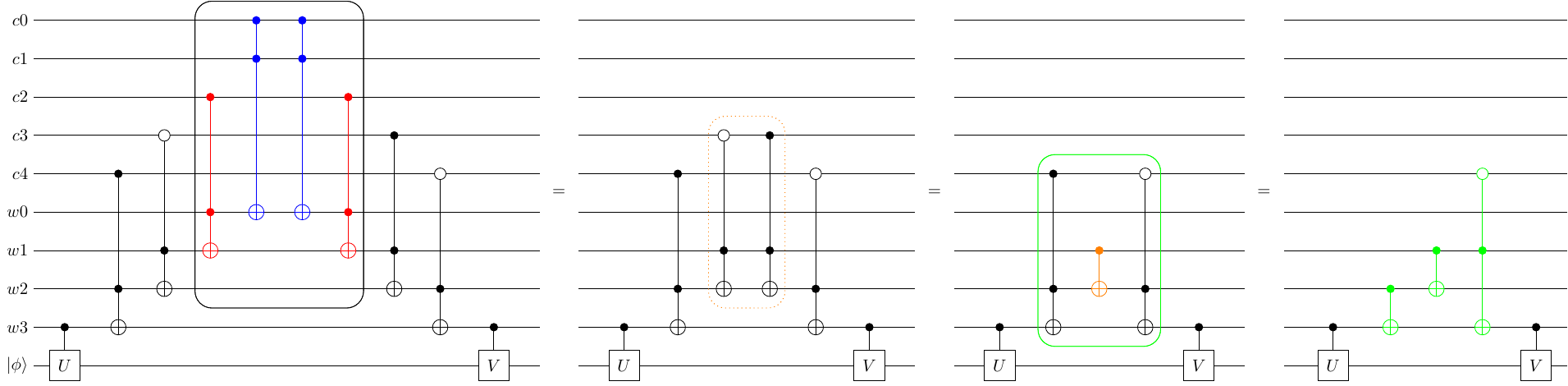}
\caption{Example reduction in binary code for $11101\text{-}U$ and $11110\text{-}V$.}
\label{fig:binary_partial}
  \end{subfigure}
  \caption{Example partial reductions when control strings between adjacent unitaries differ by $x$ leading bits.}\label{fig:partial_BG}
\end{figure}

\begin{figure}[h!]
\centering
  \begin{subfigure}[b]{1\textwidth}
\includegraphics[scale=0.5]{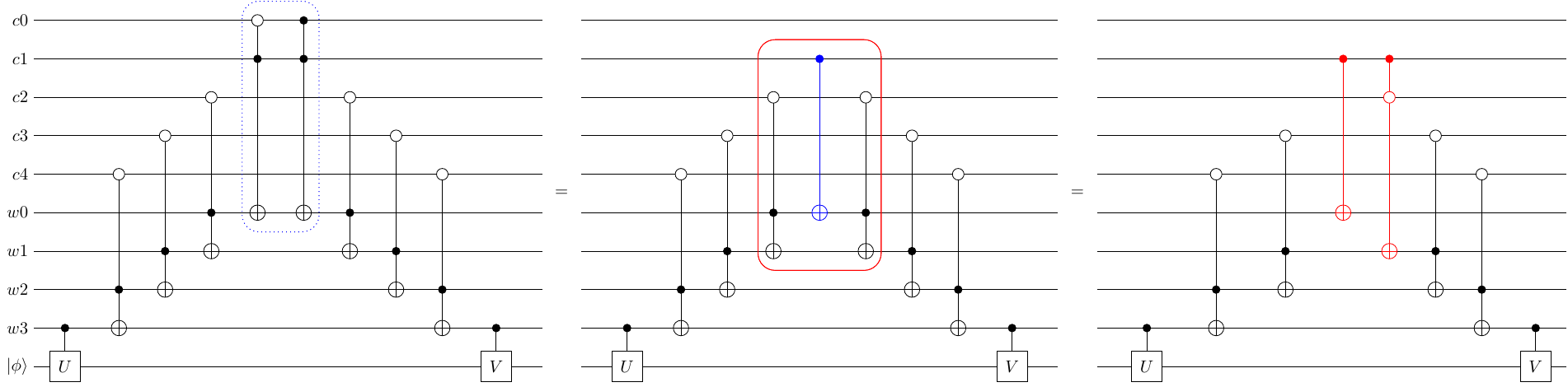}
\caption{Gray case:  $01000\text{-}U$ and $11000\text{-}V$}
\label{fig:grey_singular}
  \end{subfigure}
  \begin{subfigure}[b]{1\textwidth}
\includegraphics[scale=0.5]{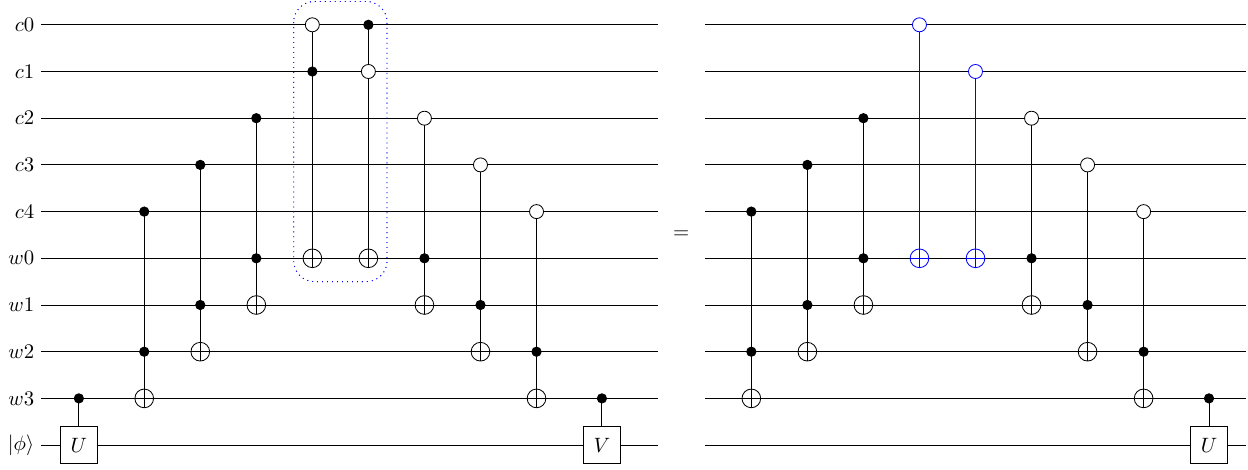}
\caption{Binary case:  $01111\text{-}U$ and $10000\text{-}V$}
\label{fig:binary_singular}
  \end{subfigure}
  \caption{Singular case when leading bit between adjacent control unitaries differ ($x=0$).}\label{fig:singular_BG}
\end{figure}

For $0 \leq x <(N_{c}-1)$, after the trivial Toffoli simplifications, it will always be possible to convert the two Toffoli gates into a CNOT gate. Again they must differ by one bit in a Gray code and the template in Fig. \ref{fig:TT_1b} can be applied. The remaining circuit will have a Toffoli-CNOT-Toffoli. These can be further reduced by applying a template shown in Fig. \ref{fig:TxT_simplifications}. In a Gray encoding, the template in Fig. \ref{fig:TxT_simplifications} can always be applied, giving an optimal reduction. Fig. \ref{fig:partial_BG}a shows an example case with a Gray encoding.  Fig. \ref{fig:partial_BG}b shows how a less optimal reduction, which occurs when a binary encoding is used. Fig. \ref{fig:singular_BG} shows another example, where a binary encoding again results in a less optimal reduction compared to a Gray code.

In a Gray encoding, $3$ Toffoli gates will be cancelled at each step and $1$ CNOT gate generated. The number of Toffoli gates removed in this process is given by: 

\begin{equation}
\begin{aligned}
\sum_{x=0}^{k} 2^{x}(3) &=  3(2^{k+1}-1)   \\
     &= 3(2^{N_{c}-1}-1)\text{ where } k=N_{c}-2 .
\end{aligned}
\end{equation}

\noindent The increase in CNOT count is given in equation \ref{eq:CNOT_increase_Toff_cancel}.

\begin{equation}
\label{eq:CNOT_increase_Toff_cancel}
\begin{aligned}
 \sum_{x=0}^{k}  2^{x} &=  2^{k+1}-1   \\
     &=2^{N_{c}-1}-1  \text{ where } k=N_{c}-2.
\end{aligned}
\end{equation}

The total number of CNOT gates is given by equation \ref{eq:CNOT_increase_Toff_cancel}. In a Gray encoding, the optimized number of Toffoli gates is given by equation \ref{eq:TOFF_scaling}. Here $T$ is short for Toffoli.

\onecolumngrid
\begin{equation}
\label{eq:TOFF_scaling}
\begin{aligned}
\text{Toffoli count} &= \underbrace{\Big(2^{N_{c}} (2N_{c}-2) \Big)}_{\text{No reductions}} - \underbrace{\Big( 2^{N_{c}+1}(N_{c}-3)+8 \Big)}_{\text{trivial TT}} \\
&- \underbrace{\Big( 2^{N_{c}} \Big)}_{\text{TT to CNOT for }x=N_{c}-1} -  \underbrace{\Big( 3(2^{N_{c}-1}-1) \Big)}_{\substack{ 0 \leq x \leq N_{c}-2 \\ \text{TT to CNOT and} \\ \text{T-CNOT-T to CNOT-T}}}  \\
&=  3(2^{N_{c}-1}) - 5.
\end{aligned}
\end{equation}

\begin{figure}[h!]
\centering
\includegraphics[width=\textwidth]{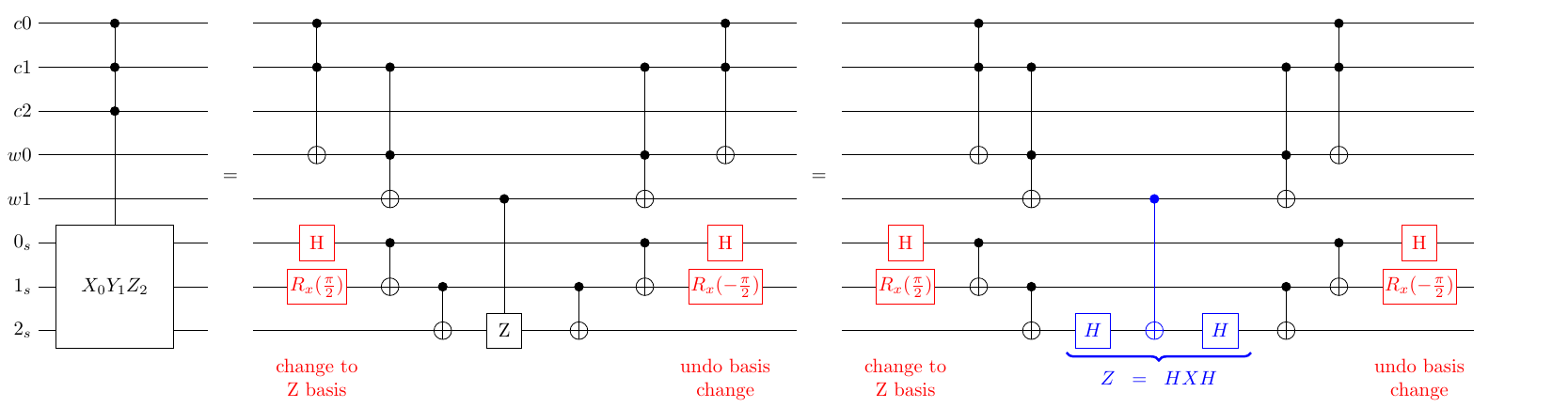}
\caption{Example quantum circuit to perform a multi-control $P_{i}$ gate via a cascade approach.}
\label{fig:LCU_cascade_example}
\end{figure}

\begin{figure}[h!]
\centering
\includegraphics[width=\textwidth]{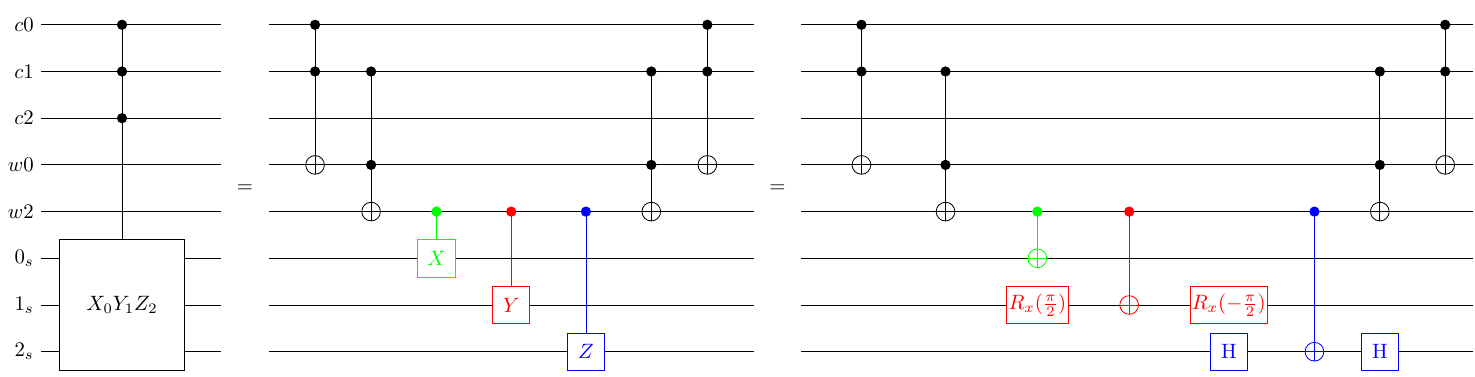}
\caption{Example quantum circuit to perform a multi-control $P_{i}$ gate via a direct approach.}
\label{fig:LCU_direct_example}
\end{figure}

\begin{figure}[h!]
\centering
\includegraphics[scale=0.9]{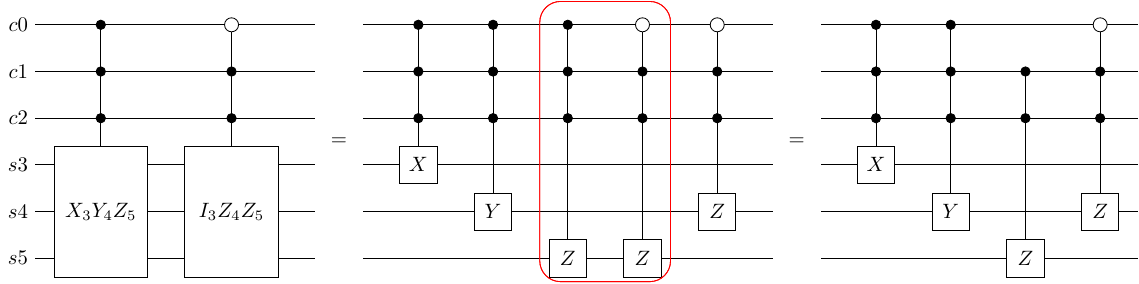}
\caption{Example lexicographical circuit simplification.}
\label{fig:LCU_lexograph}
\end{figure}

\twocolumngrid

\begin{figure*}[t]
\centering
\includegraphics[scale=0.9]{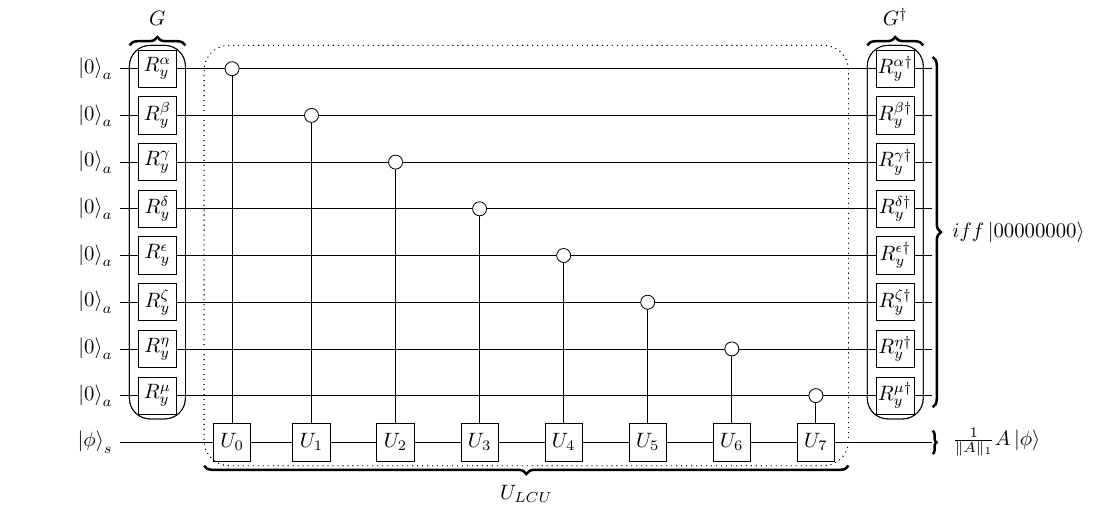}
\caption{Example unary encoded LCU circuit where $\bra{00...0}_{a}G^{\dagger}U_{LCU}G\ket{00...0}_{a} = \frac{A}{\|A\|_{1}}$.}
\label{fig:unary_circuit}
\end{figure*}

When $U_{LCU}$ is encoded using a Gray code and all $2^{N_{c}}$ control states are used there will be $3(2^{N_{c}-1})-5$ Toffoli gates and $2^{N_{c}-1}-1 $ additional CNOT gates. Each Toffoli gate can be decomposed into 9 single qubit gates and 6 CNOT gates \cite{nam2018automated}, therefore the reduced gate count requires $27(2^{N_{c}-1})-45$ single qubit gates and $19(2^{N_{c}-1})-31$ CNOT gates. 

So far these counts do not include any gate that acts on the system register, as different approaches are possible. In the next two Sections we analyse two different possibilities - a cascade and direct approach. We consider each case with a Gray encoding of the control unitaries.

\subsubsection{\label{sec:LCU_cascade} LCU cascade}
In the cascade approach, each control-$P_{i}$ operator is performed using different changes of basis, a cascade of CNOT gates and a CNOT controlled by the work qubits. This approach requires a change of basis for every $X$ and $Y$ of each $P_{i}$ gate of $U_{LCU}$ (equation \ref{eq:LCU_U_unitaryP}). In general $\mathcal{O}(2N_{s})$ gates.  The resulting operator can then be implemented using a cascade of $\mathcal{O}(2(N_{s}-1))$ CNOT gates and a control  $Z$ gate. Two Hardamard gates can convert the control $Z$ gate into a CNOT gate. An example is shown in Fig. \ref{fig:LCU_cascade_example}.

The additional gate count on the system register will scale as $\mathcal{O} \big(2^{N_{c}}(2N_{s}+2) \big)$ single qubit gates and $\mathcal{O} \big( 2^{N_{c}}(2N_{s}-1) \big)$ CNOT gates.

The full gate count in a Gray encoding, including the ancilla register, scales as $\mathcal{O} \big( 2^{N_{c}-1}(4N_{s}+31)-45 \big)$ single qubit gates and $\mathcal{O} \big( 2^{N_{c}-1}(4N_{s}+17)-31 \big)$ CNOT gates. Importantly $N_{c} = \lceil \log_{2}(|H_{S_{l}}|-1) \rceil$ and $N_{w}=N_{c}-1$. 

\subsubsection{\label{sec:LCU_control} LCU direct}

Compared to the cascade approach, the direct approach implements each control $n$-fold Pauli operator of $U_{LCU}$ (equation \ref{eq:LCU_U_unitaryP}) on the system register directly. $\mathcal{O}(2N_{s})$ change of basis gates and $\mathcal{O}(N_{s})$ CNOT gates are required per control gate. The number of single and CNOT gates scales as $\mathcal{O} \big( 2N_{s}(2^{N_{c}} \big)$ and $\mathcal{O} \big( N_{s}(2^{N_{c}}) \big)$ respectively. Fig. \ref{fig:LCU_direct_example} illustrates the approach.

The overall gate count over the system and ancilla registers scales as $\mathcal{O} \big( 2^{N_{c}-1}(4N_{s}+27)-45 \big)$ single qubit gates and $\mathcal{O} \big( 2^{N_{c}-1} (2N_{s} +19) - 31 \big)$ CNOT gates.

\subsubsection{\label{sec:LCU_special} LCU constant factor ($|H_{S_{l}}|\leq 5$)}
The scaling is different when $N_{c} \leq 2$ and no work qubits are required. For $N_{c}=1$ all gates in $U_{LCU}$ are controlled by one ancilla qubit. The circuit for $G$ (equation \ref{eq:LCU_G_unitaryP}) is defined by a single $R_{y}$ rotation. The circuit for $U_{LCU}$  is implemented by $\mathcal{O}(4(N_{s}))$ change of basis gates and $\mathcal{O}(2N_{s})$ CNOT gates.

For $N_{c}=2$, the direct and cascade approach can be used to construct $U_{LCU}$. The direct approach requires $\mathcal{O}(4N_{s})$ Toffoli and $\mathcal{O}(8N_{s})$ change of basis gates. The cascade approach requires $\mathcal{O}(8N_{s})$ changes of basis, $\mathcal{O}(8N_{s})$ CNOT and $\mathcal{O}(4)$ Toffoli gates.

By limiting the size of each anticommuting clique to $|H_{S_{l}}|\leq 5$ $\forall \{l\}_{l=0,1,...,m_{c}-1}$ no work qubits will be required to implement $R_{l}$. 

The quantum circuits required to implement unitary partitioning under these conditions are realistic for implementation on current and near term devices. This offers a constant factor improvement on the number of measurements required.

If anticommuting cliques $|H_{S_{l}}|> 5$ are present, they can be partitioned into separate subsets each of size less than 5. The produced sets will still be valid anticommuting cliques. A re-normalization on all subsets must also be performed.

\subsubsection{\label{sec:LCU_lexo} Further LCU simplifications}
An additional circuit simplification is possible cancellations in the gates making up $U_{LCU}$. We did not explicitly consider this in our work here, but note its clear application. The ordering of the control unitaries in $U_{LCU}$ is arbitrary and common qubit-wise Pauli strings can be cancelled. The optimal reduction is obtained if Pauli operators on common qubits are maximised - this is known as a lexicographical ordering. An example reduction is given in Fig. \ref{fig:LCU_lexograph}.

We did not employ this process for our \ce{H2} simulation, as it offered no improvement. The \ce{LiH} problem would have benefited from this reduction however, as we only simulated this problem on a QPU emulator multi-control gates could be performed directly. We therefore didn't decompose these operations into their single and two qubit gates and simulated the control $P_{i}$ gates directly.

\begin{table}[b]
\centering
\begin{tabular}{cccc}
\hline
Decimal & Binary & Gray & Unary    \\ \hline
0       & 000    & 000  & 00000001 \\
1       & 001    & 001  & 00000010 \\
2       & 010    & 011  & 00000100 \\
3       & 011    & 010  & 00001000 \\
4       & 100    & 110  & 00010000 \\
5       & 101    & 111  & 00100000 \\
6       & 110    & 101  & 01000000 \\
7       & 111    & 100  & 10000000 \\ \hline
\end{tabular}
\caption{Example of unary encoding scheme, compared to decimal, binary and Gray encodings.} 
\label{tab:unary_encoding}
\end{table}

\begin{table*}[t]
\centering
\begin{adjustbox}{width=0.95\textwidth}
\small
\begin{tabular}{cccccc}
\hline
Method                                                     & $N_{c}$                                                  & $N_{w}$                    & G circuit                                                                                                                                                              & Ancilla Reg $(U_{LCU})$                                                                                                                                                                  & System Reg $(U_{LCU})$                                                                                                                                                         \\ \hline
\begin{tabular}[c]{@{}c@{}}LCU Gray\\ (cascade)\end{tabular} & \multirow{2}{*}{$\lceil \log_{2}(|H_{S_{l}}|-1) \rceil$} & \multirow{2}{*}{$N_{c}-1$} & \multirow{2}{*}{\begin{tabular}[c]{@{}c@{}}$\mathcal{O}(N_{c} \: log_{2}(N_{c})^{2})$\\ standard 1 and 2-bit gate \\ operations \cite{long2001efficient}\end{tabular}} & \multirow{2}{*}{\begin{tabular}[c]{@{}c@{}}single: $\mathcal{O} \big( 2^{N_{c}-1} N_{s}\big)$\\ CNOT: $\mathcal{O} \big( 2^{N_{c}-1} N_{s} \big)$\end{tabular}} & \begin{tabular}[c]{@{}c@{}}single: $\mathcal{O} \big(2^{N_{c}}N_{s} \big)$\\ CNOT: $\mathcal{O} \big( 2^{N_{c}} N_{s} \big)$\end{tabular}                 \\ \cline{1-1} \cline{6-6} 
\begin{tabular}[c]{@{}c@{}}LCU Gray\\ (direct)\end{tabular}  &                                                          &                            &                                                                                                                                                                        &                                                                                                                                                                               & \begin{tabular}[c]{@{}c@{}}single: $\mathcal{O} \big( 2^{N_{c}}N_{s} \big)$\\ CNOT:$\mathcal{O} \big( 2^{N_{c}}N_{s} \big)$\end{tabular}                       \\ \hline
\begin{tabular}[c]{@{}c@{}}LCU unary\\ (cascade)\end{tabular}    & \multirow{2}{*}{$|H_{S_{l}}|-1$}                         & \multirow{2}{*}{N/A}       & \multirow{2}{*}{single: $\mathcal{O} \big( |H_{S_{l}}| \big)$}                                                                                                    & \multirow{2}{*}{N/A}                                                                                                                                                          & \begin{tabular}[c]{@{}c@{}}single: $\mathcal{O} \big(N_{c} N_{s} \big)$\\ CNOT: $\mathcal{O} \big( N_{c}N_{s} \big)$\end{tabular}                 \\ \cline{1-1} \cline{6-6} 
\begin{tabular}[c]{@{}c@{}}LCU unary\\ (direct)\end{tabular}     &                                                          &                            &                                                                                                                                                                        &                                                                                                                                                                               & \begin{tabular}[c]{@{}c@{}}single: $\mathcal{O} \big( N_{s}N_{c} \big)$\\ CNOT:$\mathcal{O} \big( N_{s}N_{c} \big)$\end{tabular}                       \\ \hline
SeqRot                                                      & N/A                                                      & N/A                        & N/A                                                                                                                                                                    & N/A                                                                                                                                                                           & \begin{tabular}[c]{@{}c@{}}single: $\mathcal{O} \big( N_{s} |H_{S_{l}}| \big)$\\ CNOT: $\mathcal{O} \big( N_{s} |H_{S_{l}}|  \big)$\end{tabular} \\ \hline
\end{tabular}
\end{adjustbox}
\caption{Upper bound for different resources required to perform $R_{l}$ for unitary partitioning via different implementations. LCU Gray and LCU unary represent the cases when $U_{LCU}$ (equation \ref{eq:LCU_U}) is encoded via a Gray and unary encoding respectively. $N_{c}$ and $N_{w}$ are the number of extra control and work qubits required by a given LCU implementation.} 
\label{tab:circuit_scal}
\end{table*}

\subsection{\label{sec:unary} LCU Unary Implementation}

Another approach to implementing the LCU is having each control unitary in equation \ref{eq:LCU_U_unitaryP} controlled by its own qubit. Hence the number of control qubits required is $N_{c}=|H_{S_{l}}|-1$. This is known as a unary or one-hot encoding \cite{sawaya2020resource}. An example encoding for 8 states is given in Table \ref{tab:unary_encoding}.

Under a unary encoding, the quantum circuit for $G$ (equation \ref{eq:LCU_G_unitaryP}) is made up of a two $R_{y}$ rotation on each qubit, where the amplitude $\sqrt{\frac{\alpha_{q}}{\|\vec{\alpha_{q}}\|_{1}}}$ is encoded on either the zero or one state. This is due to each $\alpha_{q}$ being real and positive. The number of single qubit gates on the ancilla register scales as $\mathcal{O} \big( 2(|H_{S_{l}}|-1) \big)$. No work qubits are required. The cascade or direct approach can then be utilized to implement the gates which act on the system register.

Fig. \ref{fig:unary_circuit} illustrates this approach with each amplitude encoded on the $\ket{0}_{a}$ state. This implementation  uses  an exponentially small subspace of the ancilla qubits' Hilbert space \cite{sawaya2020resource}. This isn't an efficient use of quantum memory, but reduces the circuit depth of LCU significantly.

\subsection{\label{sec:circ_analysis_discussion} Discussion}

Table \ref{tab:circuit_scal} summaries the different resources required to implement $R_{l}$ via different approaches.

In summary, the sequence of rotations implementation of unitary partitioning gives the lowest gate count. For the different LCU implementations, the direct unary approach provides the lowest depth quantum circuits at the cost of requiring an ancilla qubit for each unitary in $U_{LCU}^{(l)}$.

Recently, the largest implementation of VQE to date was only able to make use of $12$ qubits out of $53$ available~\cite{google2020hartree} and these unused qubits could be utilized for unary encoding. When this overhead becomes prohibitively large for sizeable $H_{S_{l}}$ sets, the Gray encoding schemes can be used. 

\section{\label{sec:Conc}Conclusion}
Our work shows that the unitary partitioning technique for measurement reduction can significantly improve the precision of variational calculations. For a fixed measurement budget $M$, fewer terms need to be estimated and thus the total number of separate energy estimates is increased. As the sample standard deviation of energies is similar for the different approaches, the standard error of the mean will be lower when unitary partitioning is applied. Our results indicate the deterministic sequence of rotations implementation offers the best improvement, which we find in our noiseless simulation of \ce{H2} and \ce{LiH}. In contrast, the LCU approach is probabilistic and some measurements must be discarded. The advantage over standard VQE is thus dependent on the success probability. This naive implementation of LCU can be improved by using oblivious and standard amplitude amplification \cite{OblivAmp14, grover1997quantum,guerreschi2019repeat, boyer1998tight}, which can boost the probability of success. However, further coherent resources are required.

The experimental results obtained using IBM's NISQ device (ibmqx2) show VQE with unitary partitioning applied performs no worse than conventional VQE for a fixed $M$, when combined with error mitigation techniques. Even though unitary partitioning requires fewer terms to be combined to give an energy estimate leading to less statistical noise and more energy samples, we suspect the additional coherent resources required causes an increased error accumulation, which offsets the advantages given by the technique. As quantum devices continue to improve, this effect should be reduced and we expect unitary partitioning will benefit many variational quantum algorithms.

Our work shows how precision can be improved for a fixed number of calls to a QPU; however, an alternate outlook is how this technique may allow larger problems to be studied.  For a given precision, applying unitary partitioning requires fewer samples and thus may allow larger scale simulations to be performed on reasonable timescales.

Future work should investigate how the variance of energies obtained changes if different terms in $H_{S_{l}}$ are reduced to, as there is flexibility in the unitary partitioning technique. We also note that this work has an interesting application to the recently proposed frugal shot Rosalin optimizer \cite{arrasmith2020operator}, which uses a weighted random sampling of Hamiltonian terms. Unitary partitioning transforms the Hamiltonian of interest into one with fewer terms of different coefficients; the effect this has on the optimizer's performance is an interesting avenue to explore.

\begin{acknowledgments}
\textit{Acknowledgments.---} A. R. acknowledges  support from the Engineering and Physical Sciences Research Council (EP/L015242/1). P.  J.  L.  and A. T. acknowledge  support  by the NSF STAQ project (PHY-1818914).  P. V. C. is grateful for funding from the European Commission for VECMA (800925). A. R. would also like to thank J. Dborin for useful discussions. 
\end{acknowledgments}

\nocite{zhao2019measurement, wiebe2012hamiltonian, Low2019hamiltonian, OblivAmp14, grover1997quantum, boyer1998tight, subramanian2019implementing, whitfield2011simulation, long2001efficient, nielsen2002quantum, rahman2014templates, babbush2018encoding, frank1953pulse, nam2018automated, mcclean2020openfermion, sun2018pyscf, o2016scalable, SciPyProceedings_11, leymann2020bitter, qiskit, efron1992bootstrap}

\bibliography{references}

\onecolumngrid
\appendix

\section{\label{sec:numerical_details_app} Numerical study details}
The ability of the unitary partitioning measurement reduction strategy is dependent on the problem Hamiltonian. To understand the performance of these methods we investigate Hamiltonians of interest in quantum chemistry. We consider Hamiltonians for \ce{H2} and \ce{LiH} molecules, each obtained using Openfermion-PySCF \cite{mcclean2020openfermion, sun2018pyscf}. These were converted into the qubit Hamiltonian using the Bravyi-Kitaev transformation in OpenFermion \cite{mcclean2020openfermion}. The following Sections give numerical details.

\subsection{\label{sec:H2}Molecular Hydrogen}

\begin{table}[t]
\centering
\begin{tabular}{cc}
\hline
 $l$ index &                                                        $H_{S_{l}}$ \\
 \hline
       0 &                                  $(0.5731061703432151+0j)Z_{0} Z_{1}$ \\
       1 &  $(0.2460355896585992+0j)I_{0} I_{1}$\\
       2 &  $\{(-0.4468630738162712+0j)I_{0}Z_{1}, (0.09060523100759853+0j)X_{0} X_{1} \}$\\
       3 &  $\{(0.3428256528955378+0j)Z_{0}I_{1}, (0.09060523100759853+0j)Y_{0} Y_{1} \}$\\
       \hline
\end{tabular}
\caption{Qubit Hamiltonian for \ce{H2} (equation \ref{eq:Hq_H2_REDUCED}) at a  bond length of $0.74$ \AA{} partitioned into anticommuting sets $H_{S_{l}}$.} 
\label{tab:H2_parition_H_q}
\end{table}

In the minimal STO-3G basis the qubit Hamiltonian for \ce{H2} in the BK representation is:

\begin{equation}
    \label{eq:Hq_H2}
\begin{aligned}
    H_{q}^{(H_{2})} = c_{0}\mathcal{I}+
                      c_{1} X_{0} Z_{1} X_{2} +
                      c_{2} X_{0} Z_{1} X_{2} Z_{3} +
                      c_{3} Y_{0} Z_{1} Y_{2} +
                      c_{4} Y_{0} Z_{1} Y_{2} Z_{3} + \\
                      c_{5} Z_{0} +
                      c_{6} Z_{0} Z_{1} +
                      c_{7} Z_{0} Z_{1} Z_{2} +
                      c_{8} Z_{0} Z_{1} Z_{2} Z_{3} +
                      c_{9} Z_{0} Z_{2} + \\
                      c_{10} Z_{0} Z_{2} Z_{3} +
                      c_{11} Z_{1}  +
                      c_{12} Z_{1} Z_{2} Z_{3} +
                      c_{13} Z_{1} Z_{3} +
                      c_{14} Z_{2}.
\end{aligned}
\end{equation}

\noindent This Hamiltonian only acts off diagonally on qubits $0$ and $2$ \cite{o2016scalable}, therefore it can be reduced to:

\begin{equation}
    \label{eq:Hq_H2_REDUCED}
\begin{aligned}
    H_{q}^{(H_{2})} = (c_{0} + c_{11}+ c_{13}) \mathcal{I}+
                      (c_{1} + c_{2}) X_{0} X_{2} +
                      (c_{3}+c_{4}) Y_{0} Y_{2}+
                      (c_{5} + c_{6}) Z_{0} + \\
                      (c_{7}+c_{8}+c_{10}) Z_{0} Z_{2} +
                      (c_{9}+c_{12}+c_{14}) Z_{2}.
\end{aligned}
\end{equation}

Overall 2 qubits are required, and any Pauli operator indexed with a $2$ is re-labelled with an index of $1$. The input state was found by diagonalising the problem Hamiltonian (Equation \ref{eq:Hq_H2_REDUCED}) - $\ket{\psi_{H_{2}}^{ground}} = -0.1125\ket{01} + 0.9936\ket{10}$. For our calculation, the bond length was set to R(H-H)=$0.74$ \AA{}. Note that we index the state from left to right. 

To perform unitary partitioning, the qubit Hamiltonian for \ce{H2} needs to be split into anticommuting sets $H_{S_{l}}$. As discussed in the main text, the NetworkX package was utilized to do this \cite{SciPyProceedings_11}. First, a graph of the qubit Hamiltonian was built, where each node is a term in the Hamiltonian. Next edges are put between nodes on the graph that anticommute. Finally, a graph colouring of the complement graph was performed. This searches for the minimum number of colours required to colour the graph, where no neighbours of a node can have the same color as the node itself. The ``largest first'' colouring strategy in NetworkX was used.  Each unique colour represents an anticommuting clique. Fig. \ref{fig:H2_Graph} shows the method applied to \ce{H2} and Table \ref{tab:H2_parition_H_q} gives the resulting anticommuting sets $H_{S_{l}}$ obtained. Note the set index $l$ represents a unique colour obtained in the graph colouring. This approach is the minimum clique cover problem mapped to a graph colouring problem.

 The following subsections give the quantum circuits used to estimate the ground state of \ce{H2} by standard VQE, and VQE with unitary partitioning applied. 

\subsubsection{\label{sec:H2_seq_rot} Sequence of rotations quantum circuits}

\begin{table}[h!]
  \centering
    \begin{tabular}{ccccc}
\hline
 $l$ index &        $\gamma_{l}$ & $P_{n}$ &       $\mathcal{X}_{nk}$ &       $\theta_{nk}$ \\
 \hline
       2 &      $0.455956044621043$   & $(1)Z_{1}$ & $(-1) X_{0} Y_{1}$ &  $2.941546221798205$ \\
       3 &      $0.35459658228639496$ & $(1) Z_{0}$ & $(1) X_{0} Y_{1}$  &  $0.25838176362668025$ \\
\hline
\end{tabular}
\caption{Given the partitioning in Table \ref{tab:H2_parition_H_q}, this table gives the operators and angles required to build $R_{S_{l}}$ (equation \ref{eqn:R_S_equation}) as a sequence of rotations.} 
\label{tab:H2_ROT}
\end{table}

\begin{figure}[h!]
     \centering
    \begin{subfigure}[t]{\textwidth}
        \raisebox{-\height}{\includegraphics[width=0.5\textwidth]{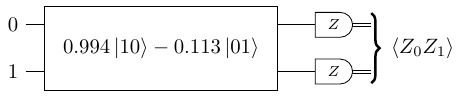}}
        \caption{$l$ index $= 0$}
    \end{subfigure}

    \begin{subfigure}[t]{\textwidth}
        \raisebox{-\height}{\includegraphics[width=\textwidth]{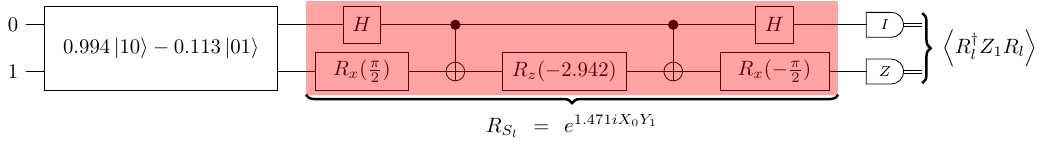}}
        \caption{$l$ index $= 2$}
    \end{subfigure}
    
    \begin{subfigure}[t]{\textwidth}
        \raisebox{-\height}{\includegraphics[width=\textwidth]{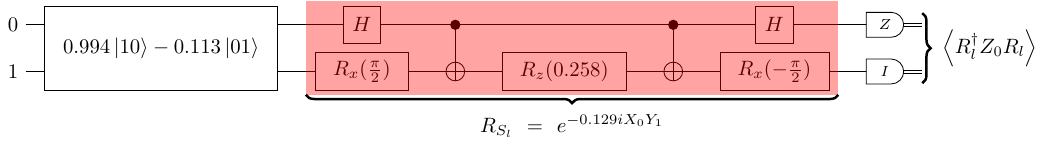}}
        \caption{$l$ index $= 3$}
    \end{subfigure}
    
    \caption{Quantum circuits required to find the ground state of \ce{H2} using VQE with unitary partitioning applied as a sequence of rotations. Tables \ref{tab:H2_parition_H_q} and \ref{tab:H2_ROT} define all operators required.}
    \label{fig:H2_seq_rot_circuits}
\end{figure}

\begin{figure}[h!]
     \centering
    \begin{subfigure}[t]{\textwidth}
        \raisebox{-\height}{\includegraphics[width=0.7\textwidth]{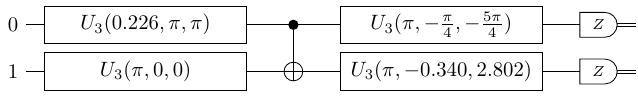}}
        \caption{$l$ index $= 0$}
    \end{subfigure}
    \begin{subfigure}[t]{\textwidth}
        \raisebox{-\height}{\includegraphics[width=\textwidth]{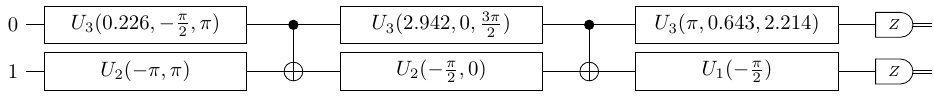}}
        \caption{$l$ index $= 2$}
    \end{subfigure}
    \begin{subfigure}[t]{\textwidth}
        \raisebox{-\height}{\includegraphics[width=\textwidth]{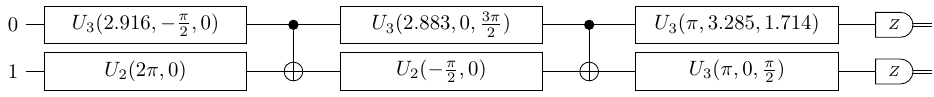}}
        \caption{$l$ index $= 3$}
    \end{subfigure}
    \caption{Quantum circuits required to find the ground state of \ce{H2} using VQE with unitary partitioning applied as a sequence of rotations. Tables \ref{tab:H2_parition_H_q} and \ref{tab:H2_ROT} define all operators required. These circuits are the compiled versions of those in Fig. \ref{fig:H2_seq_rot_circuits} for use on IBM's QPU.}
    \label{fig:H2_seq_rot_circuits_ibm}
\end{figure}

\clearpage
\subsubsection{\label{sec:H2_LCU} LCU quantum circuits}

\begin{table}[h!]
\begin{tabular}{cccc|cc}
\hline
 $l$ index    &       $\gamma_{l}$    & $P_{n}$  &  $R_{l} = \sum_{q} \alpha_{q}^{(l)} P_{q}^{(l)}$     & $U_{LCU}^{(l)} = \sum_{q} \ket{q}_{a} \bra{q}_{a} \otimes P_{q}^{(l)}$           &  $\ket{G}_{a} = \sum_{q} \sqrt{\frac{|\alpha_{q}|}{\|\alpha\|_{1}}}\ket{q}_{a} $  \\ \hline
2             &  $0.455956044621043$  & $(1)Z_{1}$  & \begin{tabular}[c]{@{}c@{}}$ (0.09985651653293749) I +$ \\ $(0.9950018472876858i) X_{0}Y_{1}$\end{tabular} & \begin{tabular}[c]{@{}c@{}}$\ket{0}_{a} \bra{0}_{a} \otimes I +$ \\ $\ket{1}_{a} \bra{1}_{a} \otimes 1i \: X_{0}Y_{1}$\end{tabular}  & \begin{tabular}[c]{@{}c@{}}$(0.3020015943219478)\ket{0}_{a}+$ \\ $(0.9533074199999713)\ket{1}_{a}$\end{tabular}  \\
3             & $0.35459658228639496$ & $(1) Z_{0}$ &\begin{tabular}[c]{@{}c@{}}\ $(0.9916664584717437) I +$ \\ $(-0.12883180951189593i) X_{0}Y_{1}$\end{tabular} & \begin{tabular}[c]{@{}c@{}}$\ket{0}_{a} \bra{0}_{a} \otimes I +$ \\ $\ket{1}_{a} \bra{1}_{a} \otimes -1i \: X_{0}Y_{1}$\end{tabular}  & \begin{tabular}[c]{@{}c@{}}$(0.9407564775082788)\ket{0}_{a}+ $\\ $(0.3390829544908076)\ket{1}_{a}$\end{tabular} \\ \hline
\end{tabular}
\caption{Given the partitioning in Table \ref{tab:H2_parition_H_q}, this table gives the operators and ancilla states required to build $R_{l}$ (equation \ref{eqn:R_LCU_re_written}) as a linear combination of unitaries.} 
\label{tab:H2_LCU}
\end{table}

\begin{figure}[h!]
     \centering
    \begin{subfigure}[t]{1\textwidth}
        \raisebox{-\height}{\includegraphics[width=0.5\textwidth]{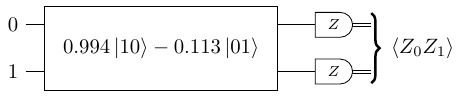}}
        \caption{$l$ index $= 0$}
    \end{subfigure}
    \begin{subfigure}[t]{\textwidth}
        \raisebox{-\height}{\includegraphics[width=0.8\textwidth]{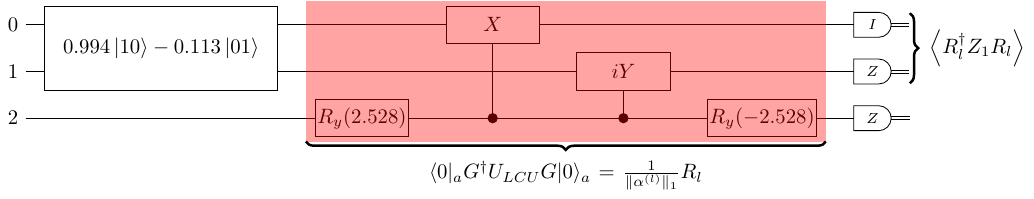}}
        \caption{$l$ index $= 2$}
    \end{subfigure}
    \begin{subfigure}[t]{\textwidth}
        \raisebox{-\height}{\includegraphics[width=0.8\textwidth]{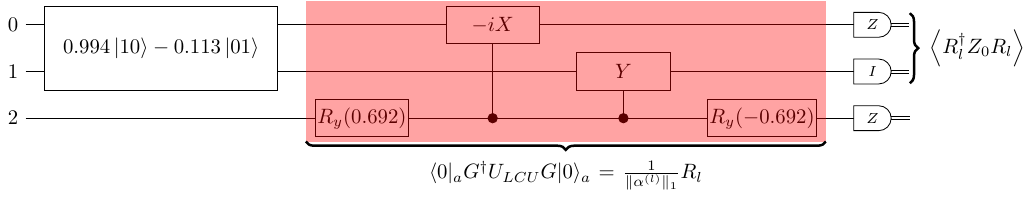}}
        \caption{$l$ index $= 3$}
    \end{subfigure}
    \caption{Quantum circuits required to find the ground state of \ce{H2} using VQE with unitary partitioning applied as a LCU. Tables \ref{tab:H2_parition_H_q} and \ref{tab:H2_LCU} define all operators and states $\ket{G}$ required.}
    \label{fig:H2_lcu_circuits}
\end{figure}

\begin{figure}[h!]
     \centering
    \begin{subfigure}[t]{\textwidth}
        \raisebox{-\height}{\includegraphics[width=0.7\textwidth]{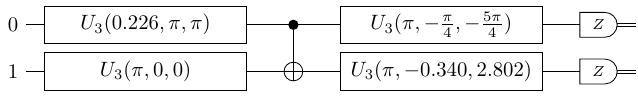}}
        \caption{$l$ index $= 0$}
    \end{subfigure}
    \begin{subfigure}[t]{\textwidth}
        \raisebox{-\height}{\includegraphics[width=\textwidth]{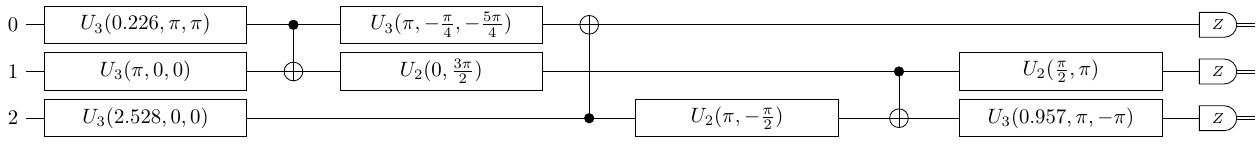}}
        \caption{$l$ index $= 2$}
    \end{subfigure}
    \begin{subfigure}[t]{\textwidth}
        \raisebox{-\height}{\includegraphics[width=\textwidth]{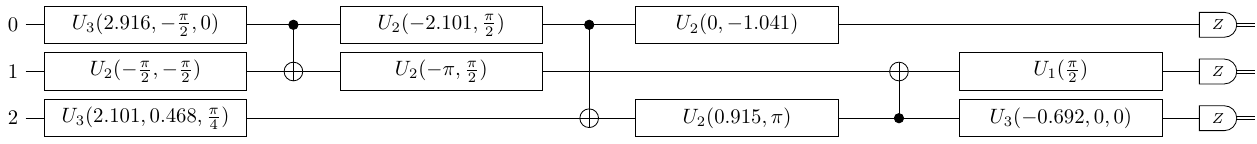}}
        \caption{$l$ index $= 3$}
    \end{subfigure}

    \caption{Quantum circuits required to find the ground state of \ce{H2} using VQE with unitary partitioning applied as a LCU. Tables \ref{tab:H2_parition_H_q} and \ref{tab:H2_LCU} define all operators and states $\ket{G}$ required. These circuits are the compiled versions of those in Fig. \ref{fig:H2_lcu_circuits} for use on IBM's QPU.}
    \label{tab:H2_lcu_circuits_ibm}
\end{figure}

\clearpage
\subsubsection{\label{sec:H2_STANDARD} Standard VQE quantum circuits}

\begin{figure}[h!]
     \centering
    \begin{subfigure}[t]{\textwidth}
        \raisebox{-\height}{\includegraphics[width=0.5\textwidth]{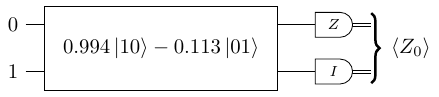}}
        \caption{}
    \end{subfigure}

    \begin{subfigure}[t]{\textwidth}
        \raisebox{-\height}{\includegraphics[width=0.5\textwidth]{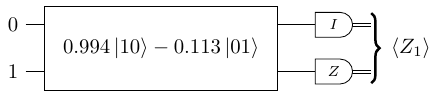}}
        \caption{}
    \end{subfigure}
    \begin{subfigure}[t]{\textwidth}
        \raisebox{-\height}{\includegraphics[width=0.6\textwidth]{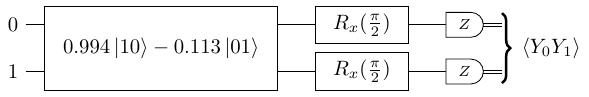}}
        \caption{}
    \end{subfigure}
    \begin{subfigure}[t]{\textwidth}
        \raisebox{-\height}{\includegraphics[width=0.55\textwidth]{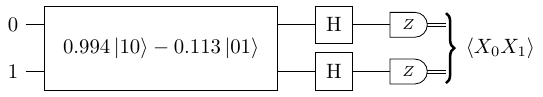}}
        \caption{}
    \end{subfigure}
    \begin{subfigure}[t]{\textwidth}
        \raisebox{-\height}{\includegraphics[width=0.5\textwidth]{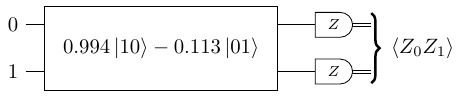}}
        \caption{}
    \end{subfigure}
    \caption{Quantum circuits required to find the ground state of \ce{H2} using standard VQE. Each operator in Table \ref{tab:H2_parition_H_q} requires a separate measurement.}
    \label{fig:H2_standard_circuits}
\end{figure}

\clearpage

\begin{figure}[t]
     \centering
    \begin{subfigure}[t]{\textwidth}
        \raisebox{-\height}{\includegraphics[width=0.7\textwidth]{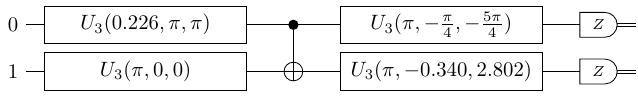}}
        \caption{}
    \end{subfigure}

    \begin{subfigure}[t]{\textwidth}
        \raisebox{-\height}{\includegraphics[width=0.7\textwidth]{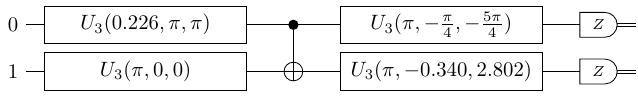}}
        \caption{}
    \end{subfigure}
    \begin{subfigure}[t]{\textwidth}
        \raisebox{-\height}{\includegraphics[width=0.7\textwidth]{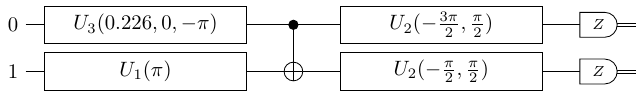}}
        \caption{}
    \end{subfigure}
    \begin{subfigure}[t]{\textwidth}
        \raisebox{-\height}{\includegraphics[width=0.7\textwidth]{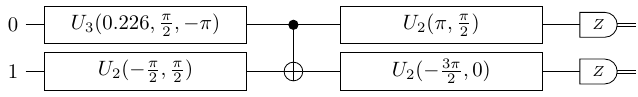}}
        \caption{}
    \end{subfigure}
    \begin{subfigure}[t]{\textwidth}
        \raisebox{-\height}{\includegraphics[width=0.7\textwidth]{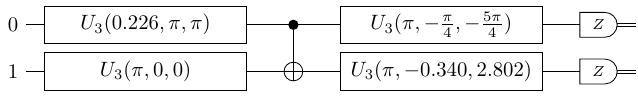}}
        \caption{}
    \end{subfigure}
    \caption{Quantum circuits required to find the ground state of \ce{H2} using standard VQE. These circuits are the compiled versions of those in Fig. \ref{fig:H2_standard_circuits} for use on IBM's QPU.}
    \label{fig:H2_standard_circuits_ibm}
\end{figure}

\newpage
\subsection{\label{sec:LiH}Lithium Hydride}

In the STO-6G basis the qubit Hamiltonian for \ce{LiH} contains $631$ terms. We considered a bond length of  R(Li-H)=$1.45$ \AA{}. In total there are $12$ spin-orbitals and the number of qubits required to simulate this system with no reductions is 12. This has been included as an XLS file.

The same method given in Section \ref{sec:H2} was used to partition the Hamiltonian into anticommuting sets and $102$ cliques were obtained. They have been included in the XLS file. The operators used to implement the sequence of rotations and LCU methods have also been included in this file.

The first sheet of the XLS document contains the \ce{LiH} Hamiltonian. The first  column contains each $c_{i}$ coefficient and the second column the associated $P_{i}$ term. In total $631$ rows. Note that $P_{i}$ operators written as $[\:]$ represent the $n$-fold identity operation.

The second sheet contains the anticommuting sets $H_{S_{l}}$. It has the same structure as Table \ref{tab:H2_parition_H_q} and contains 102 sets. 

The third sheet gives the operators required to perform unitary partitioning via a sequence of rotations, given the clique cover from the first sheet. It follows the same structure as Table \ref{tab:H2_ROT} and contains 98 entries.

The final sheet gives the operators required to perform unitary partitioning via a LCU, given the clique cover from the first sheet. It follows the same structure as Table \ref{tab:H2_LCU} and contains 98 entries.

Due to the size of each quantum circuit, they were each simulated once using qiskit's \textit{statevector\_simulator} and the final statevectors obtained were sampled from using qiskit's Statevector class \textit{sample\_counts} method \cite{qiskit}. This gives qubit measurement outcomes in the computational basis. In our simulation, each control $P_{i}$ gate required by the LCU method was directly simulated, and neither work qubits nor circuit simplifications were used.

\newpage
\subsection{\label{sec:ind_LiH_results} LiH histogram results}

\begin{figure}[h!]
     \centering
    \begin{subfigure}[t]{0.49\textwidth}
        \raisebox{-\height}{\includegraphics[width=\textwidth]{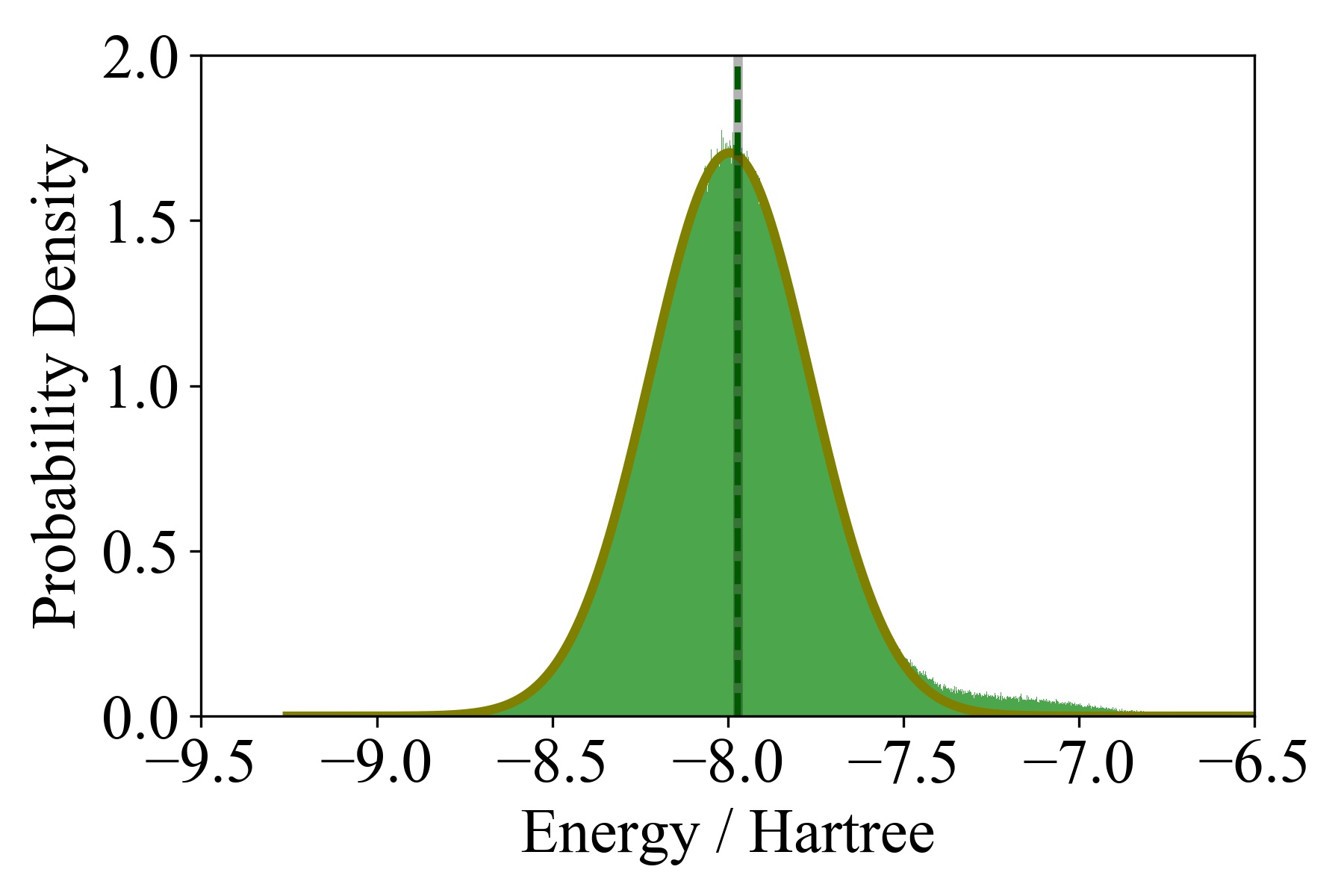}}
        \caption{}
    \end{subfigure}

    \begin{subfigure}[t]{0.49\textwidth}
        \raisebox{-\height}{\includegraphics[width=\textwidth]{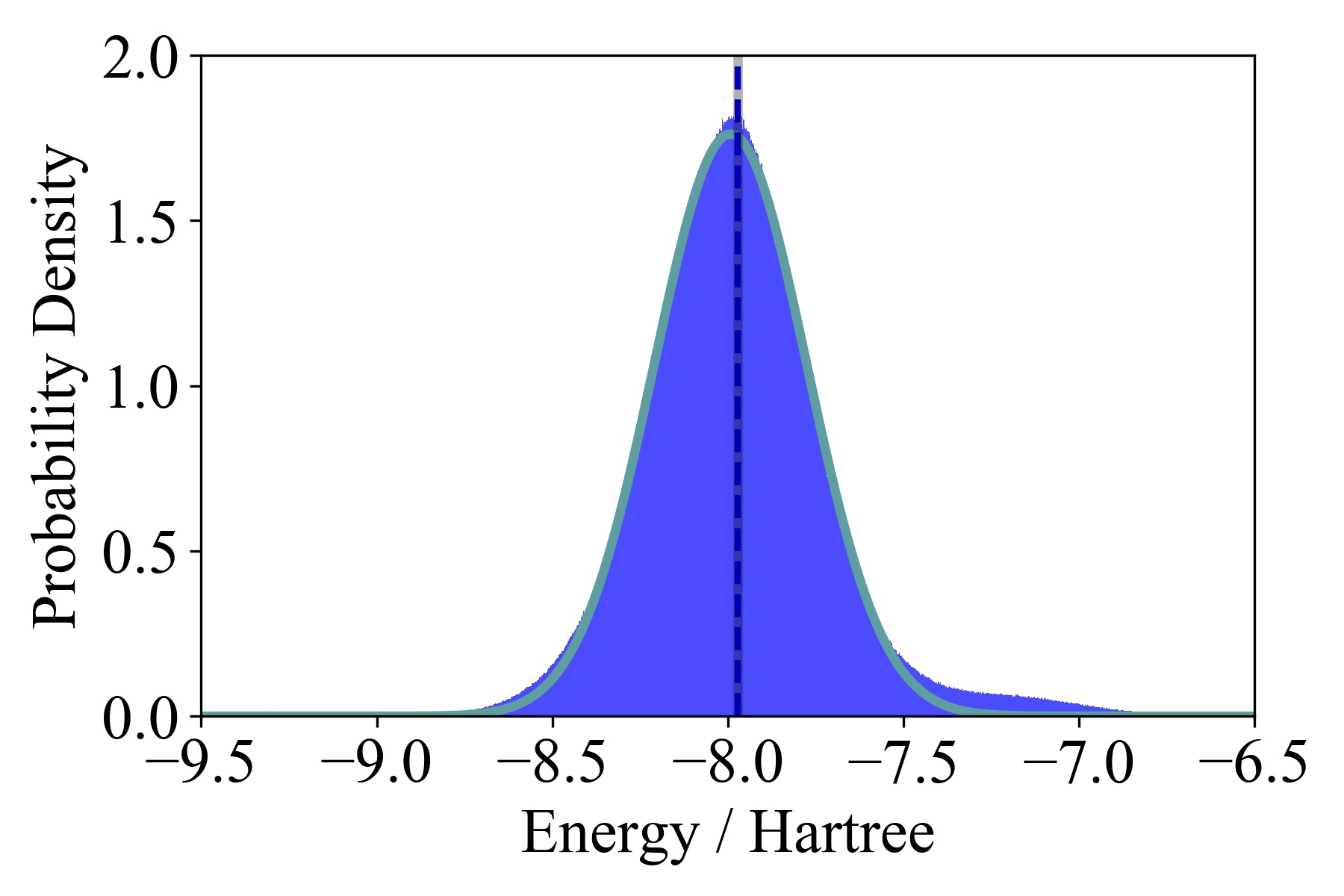}}
                \caption{}
    \end{subfigure}
    \begin{subfigure}[t]{0.49\textwidth}
        \raisebox{-\height}{\includegraphics[width=\textwidth]{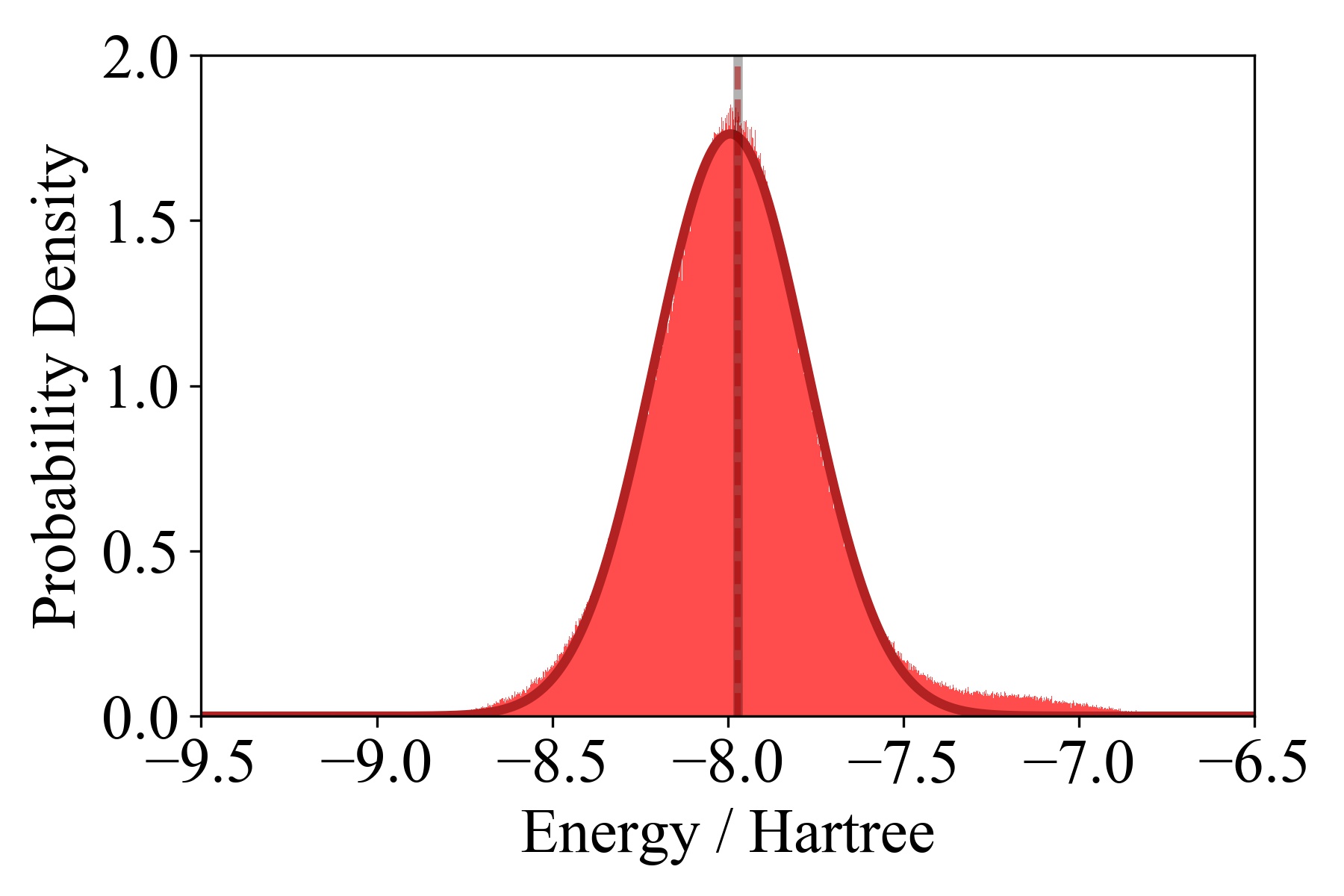}}
    \caption{}
    \end{subfigure}

\caption{Probability density function of different single-shot VQE calculations used to estimate the ground state energy $e_{j}$ of \ce{LiH}. All results here are from a noise-free QPU emulation. The green results (a) are data from standard VQE. The blue (b)  and red (c) data are results from VQE with unitary partitioning applied as a sequence of rotations and a LCU respectively. The number of bins was set to $2500$ for all histograms and a Gaussian was fitted to each result. The dashed vertical lines show the average energy for each method and each solid black vertical line shows the FCI ground state energy ($-7.97118$ $Ha)$.}
    \label{fig:individual_hist)LiH}
\end{figure}

\clearpage
\section{\label{sec:ciruit_analysis} 5-bit control example}

\begin{figure}[h!]
\centering
\includegraphics[width=0.45\textwidth]{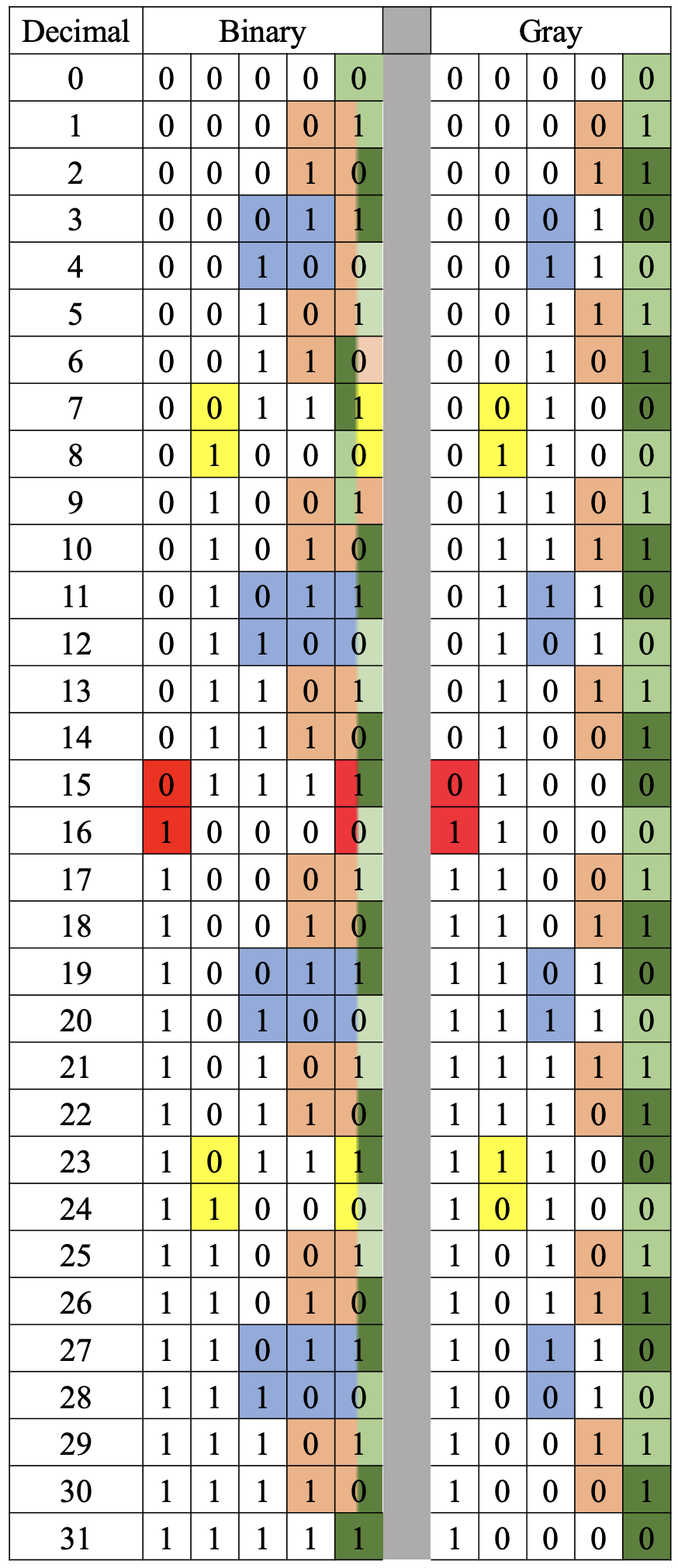}
\caption{$5$-bit Binary and Gray codes. Differences between pairs of adjacent bitstrings have been highlighted.}
\label{fig:grey_binary}
\vspace{-1.00em}
\end{figure}

\end{document}